\documentclass{aa}
\usepackage[varg]{txfonts}
\usepackage{graphicx}
\usepackage{amsmath,amstext}
\usepackage{multirow}
\bibpunct{(}{)}{;}{a}{}{,}

\begin{document}
\title{Is hemispheric asymmetry in the sunspot cycle solely governed by polar flux imbalance in the previous cycle?}

\author{Prantika Bhowmik\inst{\ref{inst1}}}
\institute{Center of Excellence in Space Sciences India, IISER Kolkata, Mohanpur 741246, West Bengal, India,\label{inst1}\\
\email{prantika.bh@gmail.com}\\
ORCID ID: \url{https://orcid.org/0000-0002-4409-7284}}

\abstract {}{Hemispheric irregularities of solar magnetic activity is a well-observed phenomenon -- the origin of which has been studied through numerical simulations and data-analysis techniques. In this work we explore possible causes generating north-south asymmetry in the reversal timing and the amplitude of polar field during cycle minimum. Additionally, we investigate how hemispheric asymmetry is translated from cycle to cycle.}{We pursue a three-step approach. Firstly, we explore the asymmetry present in the observed polar flux and sunspot area by analyzing observational data of the last 110 years. Secondly, we investigate contribution from different factors involved in the Babcock-Leighton mechanism to the evolution and generation of polar flux by performing numerical simulations with a Surface Flux Transport model and synthetic sunspot input profiles. Thirdly, translation of hemispheric asymmetry in the following cycle is estimated by assimilating simulation-generated surface magnetic field maps at cycle minimum in a dynamo simulation. Finally, we assess our understanding of hemispheric asymmetry in the context of observations by performing additional observational data-driven simulations.}{Analysis of observational data shows a profound connection between the hemispheric asymmetry in the polar flux at cycle minimum and the total hemispheric activity during the following cycle. We find that the randomness associated with the tilt angle of sunspots is the most crucial element among diverse components of the Babcock-Leighton mechanism in resulting hemispheric irregularities in the evolution of polar field. Our analyses with dynamo simulations indicate that an asymmetric poloidal field at solar minimum can introduce significant north-south asymmetry in the amplitude and timing of peak activity during the following cycle. While observational data-driven simulations reproduce salient features of the observed asymmetry in the solar cycles during the last 100 years, we speculate fluctuations in the mean field $\alpha$-effect and meridional circulation can have finite contributions in this regard.}{}

\titlerunning{Hemispheric Asymmetry in the Sunspot Cycle}
\maketitle

\section{Introduction}
The characteristics of eleven-year solar cycle are not manifested identically in the two hemispheres of the Sun. Diversity of the asymmetry between the northern and the southern hemispheres can be perceived in various observables of the solar magnetic activity -- among which sunspots are the most widely studied. Detection of hemispheric asymmetry in sunspot number goes way back to the beginning of the 20$^{th}$ century \citep{1904MNRAS..64..747M}. Utilizing long-term sunspot data series, several groups have explored various aspects of the asymmetry present in sunspot number and their associated area (\citealt{1955MNRAS.115..398N,1961says.book.....W,1977ApJS...33..391W,1986SoPh..106...35S,2001ApJ...554L.115L,2002A&A...383..648L,2005A&A...431L...5B,2006A&A...447..735T,2015LRSP...12....4H, 2016AJ....151...70D} and references therein) and established that the observed irregularities are statistically significant and cannot be achieved from a random distribution of sunspots
\citep{1993A&A...274..497C,1994SoPh..152..481O,2006A&A...447..735T}. Using historical records of sunspot data of last 300 years, \citet{2010AN....331..765Z} showed that there exits a hemispheric phase difference in the rising, peak and declining epochs associated with each cycle and the hemispheric dominance roughly changes in every eight solar cycles. The phase lag can become as high as 19 months \citep{2010SoPh..261..193N}. In a review, \citet{2014SSRv..186..251N} provided an observed upper limit of 20$\%$ of asymmetry both in the cycle amplitude and timing of peak activity. The north-south asymmetry of solar magnetic activity is also reflected in the high energetic events such as flares, coronal mass ejections (CMEs), gamma-ray and type II radio bursts (\citealt{1987SoPh..114..185V,1996SoPh..166..201A,2009MNRAS.400.1383G,1971SoPh...20..332W,2015NatCo...6E6491M} and references therein). 

Apart from the extensive observational studies on hemispheric variability of solar magnetic activity, many groups have also explored the origin of asymmetry by utilizing solar dynamo models (see reviews by \citealt{2014SSRv..186..251N} and \citealt{2015SSRv..196..101B}). The solar magnetic field is believed to be originated and sustained by a dynamo mechanism acting in the solar convection zone governed by the laws of magnetohydrodynamics (MHD) \citep{2010LRSP....7....3C}. In the framework of a solar dynamo, two components of magnetic field (known as, toroidal and poloidal components) interchange between themselves in the presence of large-scale plasma flows. While the differential rotation converts the poloidal component to the toroidal one by stretching it along the azimuthal direction \citep{1955ApJ...122..293P}, diverse conjectures exist to explain the generation of the poloidal component from the toroidal one, e.g., mean field $\alpha$-effect, MHD instabilities, hydrodynamical shear instabilities and Babcock-Leighton mechanism (see, review by \citealt{2010LRSP....7....3C}). 

Variation of the toroidal component is manifested in the modulation of sunspot number as the toroidal flux tubes satisfying the magnetic buoyancy criterion rise through the solar convection zone \citep{2009LRSP....6....4F} and emerge on the photosphere as sunspots (also known as active regions) with certain latitude-dependent tilt induced by the Coriolis force (Joy's law, \citealt{1919ApJ....49..153H}). Most of the sunspots are surfaced on the photosphere in pairs as Bipolar Magnetic Regions (BMRs) with a `leading' and a `following' spot (according to rotating direction) of opposite magnetic polarities where the leading spot predominantly appears at a lower latitude compared to the following one. For a particular cycle, if the preceding (and the succeeding) spots of the BMRs in the northern hemisphere has a positive (and negative) polarity, the leading and the following spots in the southern hemisphere will be of negative and positive polarity, respectively. But this polarity distribution will reverse during the next solar cycle -- creating a 22 years long magnetic cycle. This polarity rule is known as Hale's polarity law \citep{1925ApJ....62..270H}. 

Modulation of the poloidal component is captured in the intensity of magnetic field near the polar regions of the Sun such that polar field can serve as a proxy for quantifying the poloidal component. The importance of solar polar field which is also a measure of the global magnetic dipole of the Sun, is manifold. On one hand, it controls magnetic environment of the heliosphere by regulating the radiative and particulate output of the Sun primarily during cycle minimum. On the other, the amplitude of polar field at cycle minimum is utilized as one of the best precursors for predicting the strength of the following solar cycle \citep{2010LRSP....7....6P,2016ApJ...823L..22C,2016JGRA..12110744H}. Generation and evolution of polar field can be explained in the framework of Babcock-Leighton (B-L) mechanism \citep{1961ApJ...133..572B,1969ApJ...156....1L,1989Sci...245..712W,2010A&A...518A...7D,2010ApJ...719..264C,2014ApJ...791....5J}, where the magnetic flux associated with tilted sunspots get diffused due to turbulent diffusion (caused by turbulent motion of super-granular convective cells) and drift towards the pole aided by meridional circulation. The advected flux (primarily from the following polarities in both the hemispheres) accumulate at the poles and alter the polarity of the global solar magnetic field. This polarity reversal occurs during cycle maximum, and the polar field attains its peak value during cycle minimum. Since any significant hemispheric irregularities associated with the emerging sunspots and transport parameters involved in the B-L mechanism can influence the polar field evolution, we observe profound north-south asymmetry both in the timing of reversal as well as the final strength of polar field during cycle minimum. 

In this study, we primarily explore the origin of hemispheric asymmetry prevailed in the development and evolution of polar field by utilizing a Surface Flux Transport (SFT) model which mimics the B-L mechanism on the solar surface. Besides, we investigate the interdependency between the polar field during solar minimum and the amplitude of the following solar cycle in the context of hemispheric asymmetry by analyzing the observational data. Lastly, we study how any asymmetry present in the polar field can impact on the following solar cycle with a dynamo model while employing a hemispherically asymmetric poloidal field source at cycle minimum. We perform this analysis by assimilating surface magnetic field maps obtained from SFT simulations driven by both synthetic and observed sunspot input profiles in a continuous dynamo run and finally comparing the simulation results with actual observations. 

The paper is organized in the following fashion: in section 2, we present an analysis of the observational data followed by a brief description of the computational models utilized in our study (in section 3). In section 4, we delineate the results obtained from SFT simulations with synthetic sunspot data along with corresponding analyses. Investigation of hemispheric asymmetry with the dynamo model is detailed in section 5. Finally, we evaluate our understanding of hemispheric asymmetry in the context of solar observations in section 6. The last section (section 7) is assigned for discussions and conclusions.

\section{Hemispheric Asymmetry as Observed in the Solar Magnetic Field}

\begin{figure*}[!h]
\centering
\includegraphics[height=4.0cm,width=15.2cm]{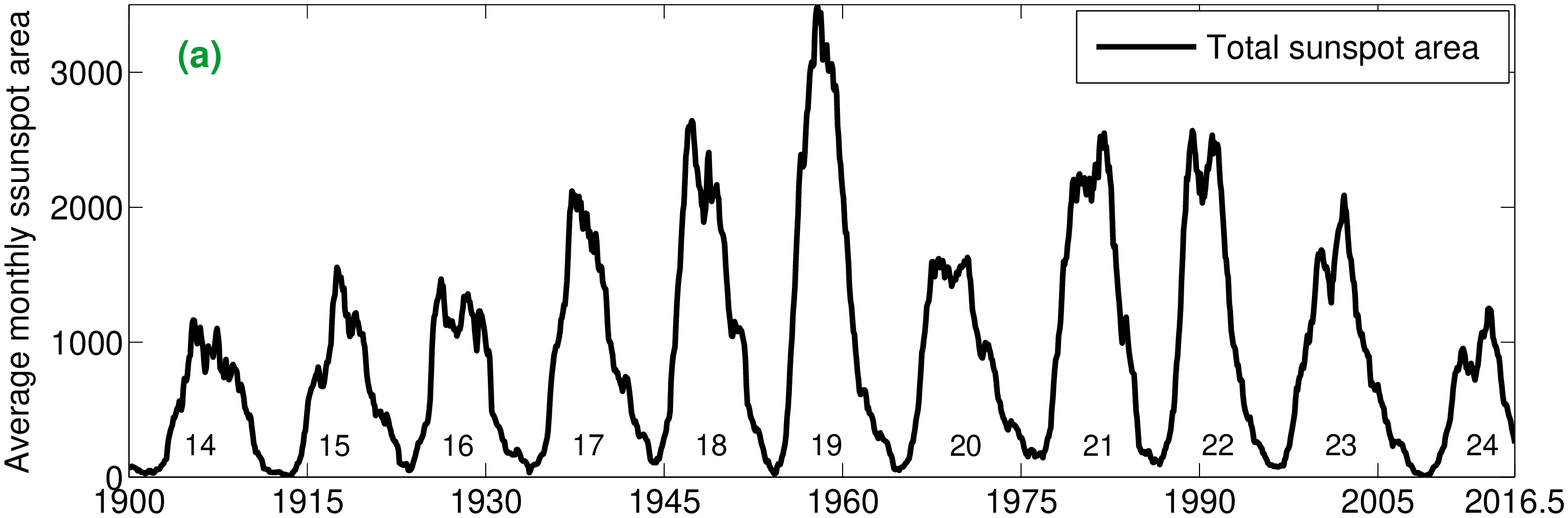}
\includegraphics[height=10.6cm, width=15.2cm]{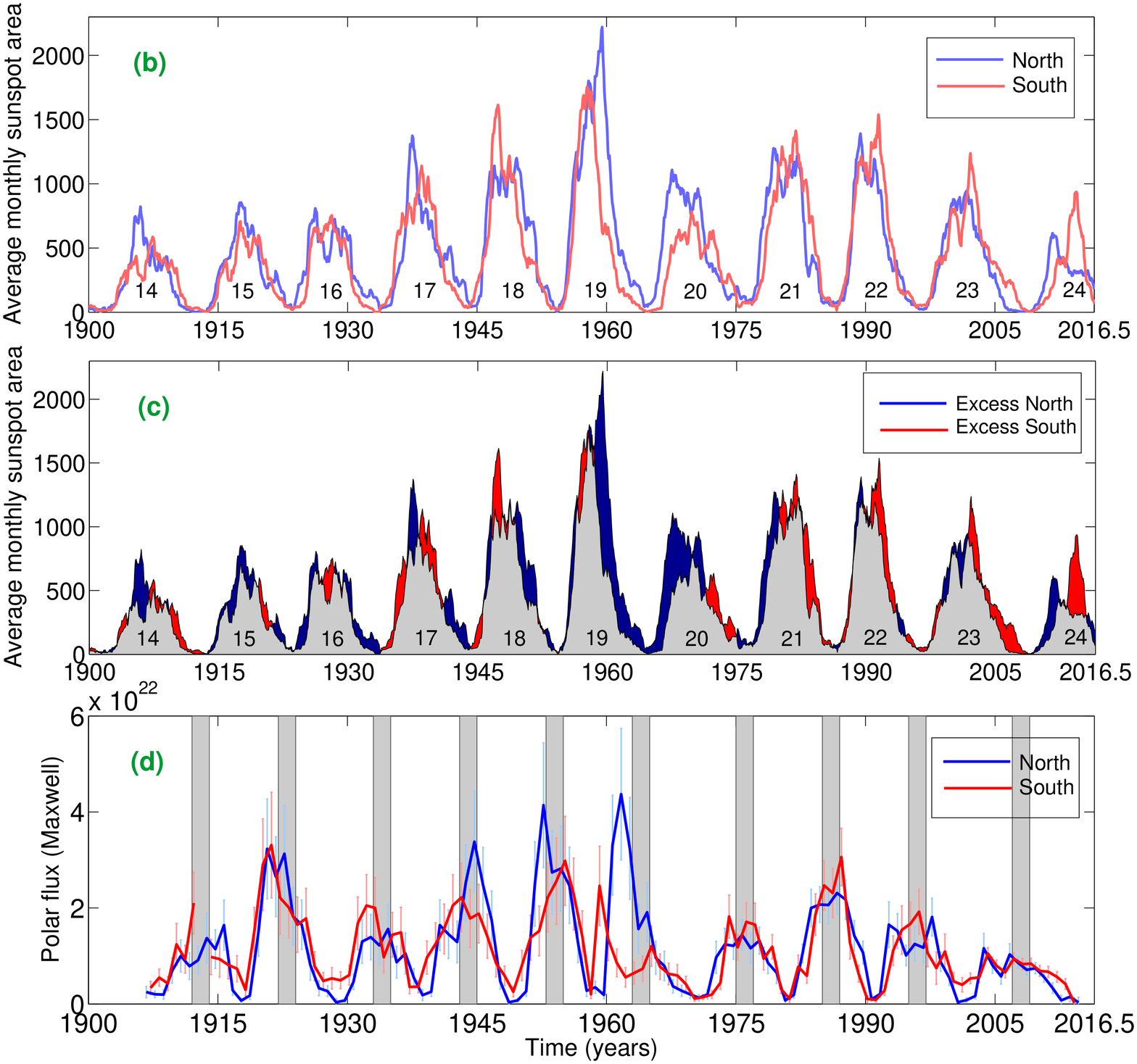}
\caption{The first panel (a) represents the time variation of averaged monthly (total) sunspot area during solar cycles 14 to 24 after performing a 13 months running average. (b) represents the same in the northern (blue curve) and the southern (red curve) hemispheres, respectively during that period. (c) depicts hemisphere-wise excess activity where the blue filled color indicates the total area associated with the north hemispheric sunspots to be greater than the southern hemispheric total sunspot area and the red filled color depicts the opposite scenario. The last panel (d) represents the variation of unsigned polar flux with error bars in two hemispheres. The rectangular gray bars depict episodes of solar minimum during 1906--2016.5 AD.} 
\label{obs_cycle}
\end{figure*}

Presence of north-south asymmetry in the monthly as well as yearly averaged sunspot area has been reported in various studies \citep{1990A&A...229..540V,1993ApJ...403..797V,1993A&A...274..497C,2002A&A...383..648L}. Here, we utilize the sunspot area data from Royal Greenwich Observatory (RGO) and USAF/NOAA database for a period spanning over 1900--2016.5 AD to investigate hemispheric asymmetry [see Fig. \ref{obs_cycle}(a)]. 

In Fig. \ref{obs_cycle}(b), we depict the 13-months running average of monthly sunspot area associated with the northern and the southern hemispheres during cycles 14--24. The figure reveals two aspects of hemispheric asymmetries which exist in each  solar cycle -- a difference in the peak amplitude as well as total sunspot area and a profound gap between the epochs of peak activity in two hemispheres. We expect this asymmetry to be similarly reflected in magnetic flux associated with sunspots as the flux is linearly proportional to the spot area \citep{2006GeoRL..33.5102D}. Fig. \ref{obs_cycle}(c) depicts the excess of monthly averaged hemispheric sunspot area for the same set of cycles -- demonstrating the relative change in cycle-phase in two hemispheres. The last panel in Fig. \ref{obs_cycle} represents time evolution of polar flux (in terms of absolute value) in the northern and the southern hemispheres during 1907--2015 AD (the associated data for polar flux is obtained from MWO calibrated polar faculae data, \citealt{2012ApJ...753..146M}). We indicate the sunspot minima by gray rectangular patches [as shown in Fig. \ref{obs_cycle}(d)], each having a width of 2 years. An earlier study by \cite{2013ApJ...767L..25M} has found polar flux at cycle minimum to be strongly correlated with the peak activity of the following sunspot cycle while utilizing the same database. 

We performed a correlation analysis between the polar flux during cycle minima (averaged over two years) and the peak sunspot area of the succeeding cycle considering two hemispheres separately. For the northern hemisphere, Pearson's linear correlation coefficient ($r_N$) is 0.62 with a p-value of 0.07, while in case of the southern hemisphere the coefficient ($r_S$) is 0.71 with a p-value of 0.03. However, excluding the data points corresponding to cycle 15 minimum and the following cycle 16 maximum from our analysis, the correlation values improve drastically. The new Pearson's correlation coefficients for two hemispheres become $r_N =$ 0.80 (p-value 0.02) and $r_S =$ 0.92 (p-value 0.001), respectively. The corresponding Spearman's rank correlation coefficients are lesser compared to the linear correlation coefficients for all datasets -- indicating the underlying mechanism connecting these two quantities to be linear in nature. While the polar flux at cycle minimum is comparable to the poloidal component of magnetic field, the sunspots and their associated area as observed on the solar surface is a manifestation of the toroidal magnetic field stored in the solar convection zone. Therefore, a high degree of linear correlation supports the theory used in various dynamo models \citep{1999ApJ...518..508D,2002Sci...296.1671N,2004A&A...427.1019C,2008ApJ...673..544Y,2014A&A...563A..18P,2014ApJ...789....5H,2016ApJ...832....9H}, where the generation of the toroidal component from the poloidal component occurs through a linear process by means of differential rotation.

\begin{figure*}[!h]
\centering
\includegraphics[height=8.2cm, width=18.0cm]{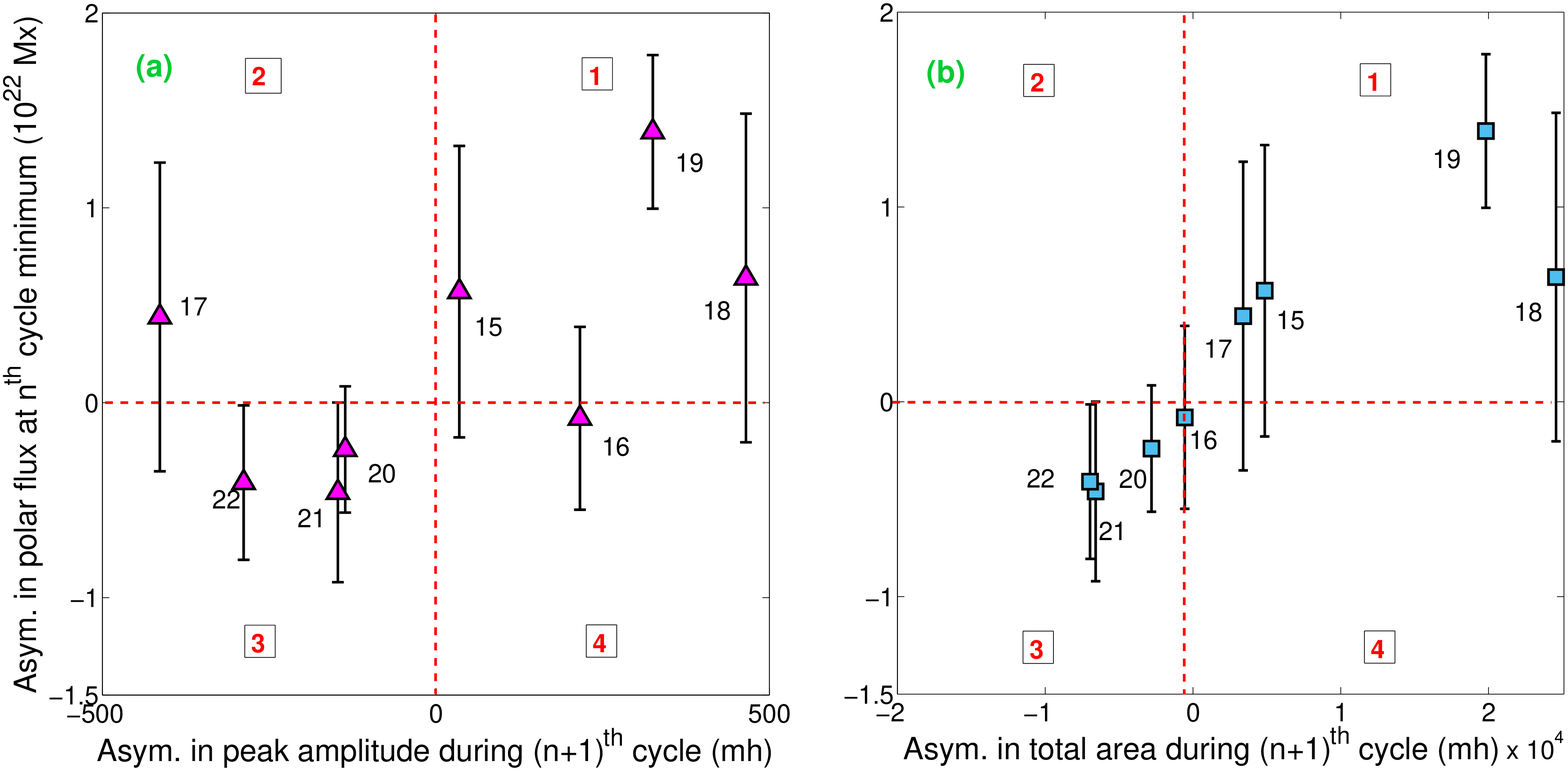}
\caption{(a) represents the hemispheric asymmetry in polar flux (Maxwells) during the minimum of the n$^{th}$ cycle (`n' varying from 15 to 22) versus the asymmetry in the sunspot area (in micro-hemispheres) during the following cycle maximum. In figure (b), the asymmetry in polar flux is compared with the asymmetry in total sunspot area of the following cycle. The numberings 1 to 4 represent the first, second, third and fourth quadrants respectively. The uncertainties present in polar flux observation are also depicted by error bars in both panels. The Spearman's rank correlation coefficient between the absolute amplitude of the asymmetry associated with the polar flux and total hemispheric sunspot area is 0.73 with p-value 0.04.} 
\label{obs_asym}
\end{figure*}

One would expect north-south asymmetry present in the polar flux during cycle minimum [see Fig. \ref{obs_cycle}(d)] to be similarly reflected in hemispheric asymmetry in the peak sunspot activity of the following cycle, such that a positive correlation should exist between them with the same hemispheric dominance \citep{2009RAA.....9..115G}. In Fig. \ref{obs_asym}(a) and (b), we compare the asymmetry in polar flux (with error bars) during n$^{th}$ cycle minimum with the asymmetry present in both the peak sunspot area and total sunspot area during (n+1)$^{th}$ cycle. Considering every possible position, all points should lie either in the first or the third quadrants to satisfy the primary requirement for positive correlation. Although this requirement is not fulfilled in case of peak sunspot area [see, Fig. \ref{obs_asym}(a)]; we find all data points satisfying the essential condition of positive correlation while considering the total sunspot area of different cycles [Fig. \ref{obs_asym}(b)]. 

A correlation analysis between the absolute values of hemispheric asymmetry in polar flux amplitude during cycle minima and the total sunspot area of the following cycle gives a Pearson linear correlation coefficient of 0.73 (with a p-value 0.04) and a Spearman rank correlation coefficient of 0.81 (with p-value 0.02). This particular result distinctly indicates that the north-south asymmetry present in polar flux (during cycle minimum) is transmitted in a non-linear manner to the asymmetry in the total sunspot area (or flux) of the following cycle. 

A careful inspection of Fig. \ref{obs_cycle}(d) reveals that hemispheric polar flux can acquire its maximum amplitude during the descending phase of a cycle and eventually settles at a comparatively lower amplitude at solar minimum. For example, during the descending phase of cycle 19, we observe that the amplitude of both the northern and the southern hemispheric polar flux becomes maximum at 1962 AD and 1959 AD (respectively), much earlier than the cycle 19 minimum (1964 AD). We denote such time instances as $t_p^{N}$ and $t_p^{S}$ for the northern and southern hemispheres, respectively. We perform a thorough correlation analysis to explore how this feature of polar flux is connected with the activity in the following cycle. We find no correlation between ($t_p^{N} - t_p^{S}$) during the n$^{th}$ cycle and the phase-lag in the hemispheric peak activity during the (n+1)$^{th}$ cycle. This result implies that a scenario where the northern hemispheric polar flux attains its maximum amplitude earlier than the south during a solar cycle (i.e., $t_p^{N}$ $<$ $t_p^{S}$) does not lead to a faster rise of the northern hemispheric activity compared to the south in the following cycle. Moreover, we carry out another correlation analysis between the amplitude of polar flux at $t_p^{N}$ and $t_p^{S}$ and the corresponding hemispheric peak activity in the (n+1)$^{th}$ cycle. We find the degree of correlation associated with both the hemispheres to be poorer ($r\approx 0.3$ on average) on contrary to the results we obtain while using the amplitude of polar flux at cycle minimum ($r\approx 0.66$ on average).

We additionally perform every correlation analysis discussed above while considering sunspot numbers instead of sunspot area. We find a similar positive linear correlation existing between the hemispheric polar flux at cycle minimum and peak amplitude of sunspot number in the following cycle -- which is quite expected as the peak sunspot number and the peak sunspot area (or flux) are well correlated (Pearson's correlation coefficient is 0.76 with a p-value of 0.01 based on last 110 years' data). The Pearson correlation coefficients associated with the northern and southern hemispheres are $r_N = 0.66$ (with a p-value of 0.053) and $r_S = 0.66$ (with a p-value of 0.054), respectively. Exclusion of polar flux at cycle 15 minimum from our analysis increases the values of correlation coefficients, such that, $r_N$ and $r_S$ become 0.78 ($p=0.02$) and 0.73 ($p=0.03$) respectively. However, the overall degree of positive linear correlation is higher in case of sunspot area compared to sunspot numbers, as the previous one is a better representative of magnetic activity of the Sun.  

In the following sections, we perform multiple computational simulations to explore probable causes instigating hemispheric asymmetry. 

\section{Computational Models}

In this work we use two disparate 2-D numerical models -- a Surface Flux Transport (SFT) model for studying the dissipation and advection of magnetic field associated with the tilted BMRs on the solar surface in the presence of magnetic diffusion and large-scale velocity fields -- the mechanism responsible for generating the polar field and a dynamo model studying the generation of toroidal field from the poloidal field in the solar convection zone. In the following section, we briefly outline these two computational models. 

\subsection{Surface Flux Transport (SFT) Model}
\subsubsection{Basic Equation}
We have developed an SFT model to study the evolution of photospheric magnetic field on the solar surface which is governed by magnetic induction equation,

\begin{equation}
\frac{\partial \mathbf{B}}{\partial t} ={\nabla} \times (\mathbf{v} \times \mathbf{B}) + \eta \nabla^{2} \mathbf{B}
\end{equation}

\noindent Where $\mathbf{v}$ represents the large scale velocities, i.e. meridional circulation and differential rotation present on the solar surface and the parameter $\eta$ is the magnetic diffusivity. As observations \citep{1993SSRv...63....1S} have shown that the surface magnetic field is predominantly along the radial direction, we numerically solve only the radial component of the induction equation which is expressed in spherical polar coordinates as,

\begin{multline}
\frac{\partial B_r}{\partial t} = -  \omega(\theta)\frac{\partial B_r}{\partial \phi} - \frac{1}{R_\odot \sin \theta} \frac{\partial}{\partial \theta}\bigg(v(\theta)B_r \sin \theta \bigg)\\
+\frac{\eta_h}{R_\odot^2}\bigg[\frac{1}{\sin \theta} \frac{\partial}{\partial \theta}\bigg(\sin \theta \frac{\partial B_r}{\partial \theta}\bigg) + \frac{1}{\sin \theta ^2}\frac{\partial ^ 2 B_r}{\partial \phi ^2}\bigg] + S(\theta, \phi, t)
\end{multline}

\noindent Where $B_r(\theta, \phi,t)$ is the radial component of magnetic field as a function of co-latitude ($\theta$) and longitude ($\phi$), $R_\odot$ is the solar radius. The axisymmetric differential rotation and meridional circulation are incorporated through $\omega(\theta)$ and $v(\theta)$ respectively. The parameter $\eta_h$ is the effective diffusion coefficient and $ S(\theta, \phi, t)$ is the source term representing the emergence of new sunspots. Since we are studying the evolution of $B_r$ on the surface of a sphere, the model has been developed using spherical harmonics. The same model was utilized earlier by \citet{2018ApJ...853...72N} for extrapolating coronal magnetic fields from surface magnetic field maps generated by SFT simulations.

\subsubsection{Transport Parameters}
The Sun has a large-scale axisymmetric rotational velocity of differential nature, i.e. plasma at different layers rotate with different speeds. This variation in velocity is observed both in radial and latitudinal direction. On the surface, the equator rotates faster than the poles. An empirical profile \citep{1983ApJ...270..288S} can express the surface differential rotation as a function of co-latitudes, 

\begin{equation}
\omega(\theta) = 13.38 - 2.30 \cos^{2}\theta - 1.62 \cos^{4}\theta  
\end{equation}

\noindent Wherein, $\omega(\theta)$ has units in degrees per day. This profile has also been validated by helioseismic observations \citep{1998ApJ...505..390S}. Another significant large-scale flow active on the solar surface is the meridional circulation which carries magnetized plasma from the equatorial region to the polar regions in both the hemispheres. The flow speed becomes zero at the equator and the poles and attains its peak amplitude near mid-latitudes. To replicate this flow in our model, we use a velocity profile prescribed by van Ballegooijen \citep{1998ApJ...501..866V}, 
\begin{equation}
 v(\lambda) =
  \begin{cases}
    - v_{0} \hspace{0.1cm \sin( \pi \lambda/ \lambda_0)} & \text{if } |\lambda | < \lambda_0  \\
   0       & \text{otherwise } 
  \end{cases}
\end{equation}\\

\noindent Where $\lambda$ is latitude in degrees ($\lambda = \pi/2 - \theta$) and $\lambda_0$ is the latitude beyond which circulation speed becomes zero which in our model is set at $\pm$75$^{\circ}$. The parameter, $v_0$ represents the maximum speed attained by the meridional circulation near mid-latitudes which varies within a range of 10--20 ms$^{-1}$. For the standard simulation, we consider the antisymmetric (about the equator) meridional circulation profile to be identical in two hemispheres with $v_0 =$ 15 ms$^{-1}$. The last transport parameter present in our model is magnetic diffusivity. It arises due to the random motions of super-granular convective cells present in the solar convection zone. We have used a constant diffusion coefficient of 250 km$^2$s$^{-1}$ which lies within the range inferred from observations \citep{2000ssma.book.....S}.

\subsubsection{Synthetic Input Profiles: Emergence of Sunspots}

The number of sunspots appearing on the solar surface roughly follows a 11-year cycle. In the beginning of the cycle, sunspots in general emerge at higher latitudes (near $\pm$ 40$^0$) and as cycle advances in time, they appear closer to the equator. This equator-ward propagation of the spots forms a structure similar to the wings of a butterfly about the equator. We consider each active region (or sunspot) associated with the synthetic input profiles as ideal BMRs with their latitudinal distribution motivated by actual observation \citep{2011A&A...528A..82J}. All active regions follow Hale's polarity law \citep{1925ApJ....62..270H} and have latitude-dependent tilt angles determined by Joy's law \citep{1919ApJ....49..153H} such that tilt angles increase with increasing latitudes. We implement Joy's law in our model by utilizing a relation, $\alpha = g \sqrt{|{\lambda}|}$, where $\alpha$ is the tilt angle and $\lambda$ is the latitudinal position of the centroid of the whole BMR \citep{2011A&A...528A..82J}. Apart from the large-scale flows, there exist localized inflows towards active regions which reduce the latitudinal separation between opposite polarities and allow a lesser amount of flux to reach the polar regions. To mimic these inflows, we introduce an additional factor $g$ in the tilt angle calculation \citep{2010ApJ...719..264C}. We choose g to be 0.7.

The active regions are randomly distributed over the full 360$^{\circ}$ range of longitude. Typically, the number of sunspots and their corresponding area in a certain solar cycle follow a power law distribution \citep{2011A&A...528A..82J}, which ensures the presence of very few sunspots with a large area. The magnetic flux associated with an active region is decided based on an empirical relation \citep{2006GeoRL..33.5102D}: $\Phi(A) = 7.0 \times 10 ^{19} A$ Maxwells, where $A$ is the area of whole sunspot in units of micro-hemispheres. The flux is equally distributed among the two polarities of the BMR. We assume the separation between the centroids of two polarities to be proportional to the spot radius obtained from its associated area. The magnetic field distribution within a single polarity follows a Gaussian distribution where the peak of the Gaussian is determined by following a prescription by \cite{1998ApJ...501..866V}. The total amount of magnetic flux associated with the sunspots of a particular synthetic input profile is about 5.4$\times$10$^{24}$ Maxwells with 3100 active regions equally distributed between two hemispheres.

\subsubsection{Initial Field Configuration}
We initialize our simulations with an ideal dipole with magnetic field primarily concentrated near the polar cap region ($\pm 70^{\circ}--90^{\circ}$) in each hemisphere. The strength (absolute value) of the polar field in each hemisphere is about 4.2 Gauss.

\subsubsection{Numerical Modeling Parameters}
Ideally, one should consider all possible values of degree ($l$) associated with spherical harmonics. Instead of taking the whole range of values of $l$ from 0 to $\infty$, we consider $l$ values varying from 0 to 63 which can spatially resolve elements with an equivalent size of supergranular cells (roughly 30 Mm) on the solar photosphere. Our SFT model is accurate up to second order in space and first order in time.

\subsection{The Solar Dynamo Model}
Recent two-dimensional kinematic solar dynamo models consider different mechanisms for generation of the poloidal component (B$_{P}$) from the toroidal component (B$_{\phi}$) of magnetic field. While majority of dynamo models \citep{1999ApJ...518..508D,2002Sci...296.1671N,2004A&A...427.1019C,2008ApJ...673..544Y} identify B-L mechanism as the sole process, others \citep{2014A&A...563A..18P,2014ApJ...789....5H} found that an additional mean field $\alpha$-effect also to be essential for sustenance of the solar dynamo. In our study, we primarily focus on generation of the toroidal component from a given poloidal component and utilize an existing 2D dynamo model where the poloidal field source term depends on both the above processes \citep{2014A&A...563A..18P}. The same model has provided satisfactory results previously \citep{2014A&A...563A..18P}. Additionally, assimilating output from the observational data-driven SFT simulations in the same dynamo model was quite successful in reproducing solar activities during the past eight solar cycles \citep{2018NatComm...9...5209B}. The basic equations used in our axisymmetric kinematic dynamo model are as follows,

\begin{equation}
    \frac{\partial A}{\partial t} + \frac{1}{s}\left( \textbf{v}_p \cdot \nabla\right)  (sA) = \eta_p\left( \nabla^2 - \frac{1}{s^2}  \right)A + \alpha B,
    \label{pol}
\end{equation}

\begin{multline}
    \frac{\partial B}{\partial t}  + s\left[ \textbf{v}_p \cdot \nabla\left(\frac{B}{s} \right) \right]
    + (\nabla \cdot \textbf{v}_p)B = \eta_t\left( \nabla^2 - \frac{1}{s^2}  \right)B \\ 
    + s\left(\left[ \nabla \times (A (r,\theta)\bf \hat{e}_\phi) \right]\cdot \nabla \Omega\right)
    + \frac{1}{s}\frac{\partial (sB)}{\partial r}\frac{\partial \eta_t}{\partial r},~~~
\end{multline}

\noindent where, $B (r, \theta)$ (i.e., $B_\phi$) and $A (r, \theta)$ are the toroidal and the poloidal (in the form of vector potential) components of magnetic field respectively. The symbols ${\bf v}_p$ and $\Omega$ is the meridional circulation and the differential rotation in the solar convection zone, and $s = r\sin(\theta)$. Two different diffusivity profiles, $\eta_t$ and $\eta_p$, are used for the toroidal and the poloidal components of magnetic field, respectively. In equation (\ref{pol}), `$\alpha B$' is the source term for generating the poloidal field which considers contribution from both the B-L mechanism and mean field $\alpha$-effect. The details of every profile and parameter used in this dynamo model are described in an already published work by \cite{2014A&A...563A..18P}.

Additional to the large-scale meridional circulation, in our dynamo model we consider advection due to turbulent pumping which is effective on the poloidal component only. The presence of downward magnetic pumping was suggested by several theoretical studies and simulations of local magneto-convection \citep{1993A&A...274..543P,1998ApJ...502L.177T,2009A&A...500..633K,1993A&A...274..543P,1996JFM...306..325B}. We assume that the much stronger component of magnetic field will remain unaffected by this downward pumping \citep{2006A&A...455..401K,2002A&A...394..735O}. Besides, \cite{2012A&A...542A.127C, 2016ApJ...832...94K} have demonstrated the importance of turbulent pumping in flux transport dynamo models in the context of compatibility with surface observations as well as SFT simulations. Thus, we add a downward radial pumping velocity with the meridional circulation ($\mathbf{v_p}$) in equation (\ref{pol}) such that the radial component of velocity ($v_r$) is changed to $v_r+\gamma_r$. The profile of the radial magnetic pumping ($\gamma_r$) is same as used in \citealt{2012ApJ...761L..13K} and given by,
\begin{multline}
    \gamma_r = -f \gamma_{0r} \Bigg[1 + erf \bigg( \frac{r - 0.715}{0.015}\bigg)\Bigg] \Bigg[1 - erf \bigg( \frac{r - 0.97}{0.1}\bigg)\Bigg] \\
    \times \Bigg[exp \bigg( \frac{r - 0.715}{0.25}\bigg)^2 \cos \theta + 1 \Bigg]
    \label{eqn_pumping}
\end{multline}
    
Amplitude of the pumping speed is controlled by the value of $f \gamma_{0r}$ and is taken as $3.6$ ms$^{-1}$. It ensures a magnetic Reynolds number of approximately $5$ \cite{2012A&A...542A.127C} and a positive dynamo growth rate where advection time due to pumping is at least five times the diffusion time. Considering the average (over the convection zone) magnetic diffusivity associated with the poloidal magnetic field to be $1.5$ km$^2$s$^{-1}$ and the width of the layer throughout which the pumping is functioning as $0.3 R_{\odot}$, $f \gamma_{0r}=3.6$ ms$^{-1}$ satisfies the mentioned conditions.

The magnetic buoyancy algorithm \citep{2002Sci...296.1671N,2004A&A...427.1019C,2014A&A...563A..18P} in dynamo simulation produces a quantity (say, B$^{Dyn}$) proportional to the strength of the toroidal field at the base of the convection zone and represents the sunspots which emerge on the solar surface after satisfying the magnetic buoyancy criterion. We utilize this proxy as a representative of sunspot cycle to study the asymmetry translated into the hemispheric activity in the following cycle from the previous cycle poloidal field.

\section{Factors Inducing Hemispheric Asymmetry in the Polar Field: An SFT Perspective}
Utilizing the SFT model described in the previous section, we investigate various sources that contribute to the north-south asymmetry in the final strength of polar field at cycle minimum. Two aspects of the B-L mechanism govern the evolution of photospheric magnetic field as well as polar field; firstly, the transport parameters on the solar surface and secondly, diverse characteristics of the emerged sunspots. Therefore, variation of these factors can originate hemispheric asymmetry in the final polar field strength, assuming the initial strength of the polar field at the beginning of cycle in two hemispheres to be precisely equal. We identify five possible sources and thoroughly investigate their effects on the final amplitude of polar field in both hemispheres. The same analysis also sheds light on the causes of relative time differences in the reversal of hemispheric polar field. 

From the time series data of observed sunspot area during last 110 years [see, Fig. \ref{obs_cycle}(b) and (c)], we can perceive three distinguishable characteristics of north-south asymmetry -- differences in (1) amplitude of peak activity, (2) timing of peak activity and (3) the total sunspot area associated with two hemispheres in a particular cycle. Following the understanding gleaned from the analysis of observational data (as described in section 2), in our current study, we give more importance to hemispheric asymmetry present in the total sunspot associated area (or flux) rather than the amplitude of peak activity. (4) Another significant source of hemispheric irregularities related to sunspot data is the presence of randomness in tilt angle of individual sunspots in addition to their systematic tilts determined by Joy's law \citep{2010A&A...518A...7D,1999SoPh..189...69S,2013SoPh..287..215M}. (5) The last hemispheric inequality incorporated in our analysis is regarding the transport parameters involved in the B-L mechanism -- application of different meridional circulation and differential rotation profiles in the northern and southern hemispheres. However, magnetic diffusivity can be assumed as homogeneous and isotropic since it is originated from the random motion of granules and super-granules on the solar surface which do not possess any directional preferences \citep{2006JGRA..11112S01N}. Thus it allows us to use a constant diffusion coefficient ($\eta_h$) in the SFT simulations. 

We explore the effect of these factors on the final amplitude of polar field through studying the time evolution of polar flux that is calculated by integrating the radial magnetic field within the polar cap regions (extended from $\pm$70$^o$ to $\pm$90$^o$ latitudes in both the hemispheres).

\begin{equation}
\Phi_p^{N/S} (t) = \int_{0^{\circ}}^{360^{\circ}}\int_{\pm 70^{\circ}}^{\pm 90^{\circ}}B_r(R_{\odot},\lambda,\phi,t) \cos\lambda d\lambda d\phi
\label{polarNS}
\end{equation}

\noindent Where $\lambda$ is latitude and $\phi$ is longitude. Besides polar flux, we also calculate the unsigned magnetic flux associated with sunspots emerging in the northern and the southern hemispheres separately. Assuming, n$^{N}_k$ number of individual spots with corresponding area, $A^N_i$ $(i=1,...,n^N_k)$ appeared on the northern hemisphere during the k$^{th}$ month, then the total sunspot-associated unsigned flux in that hemisphere during that month would be,
\begin{equation}
\Phi^{N}_{k} = \sum_{i=1}^{n^N_{k}} \Phi^N_i(A^N_i)
\end{equation}
\noindent Wherein, $\Phi^N_i(A^N_i) = 7.0 \times 10 ^{19} A^N_i$ Maxwells, while $A^N_i$ is in micro-hemispheres. Unsigned magnetic flux in the southern hemisphere is calculated following the same method. We measure the difference between $\Phi_p^{N}$ and $\Phi_p^{S}$ to quantify the asymmetry in the amplitude of hemispheric polar flux, while the difference between $\Phi^{N}_{k}$ and $\Phi^{S}_{k}$ estimates the asymmetry in hemispheric activity regarding sunspot area. 

In the following section, we separately consider each factor associated with the B-L mechanism through which hemispheric asymmetry can be introduced while keeping the other factors unaltered and investigate their individual effect on the final strength of the polar flux.

\subsection{Time Difference in Peak Activity}
For a particular cycle, one hemisphere may reach peak activity before the other, thus introducing an asymmetry. As perceived from Fig. \ref{obs_cycle},in case of 80$\%$ cycles, two hemispheres were out of phase at cycle maximum during last 110 years. For example, during solar cycle 18, the southern hemisphere reached its peak activity almost two years before the northern hemisphere reached its maximum. A similar instance was observed during cycle 19. In case of solar cycles 22 and 23, the north hemisphere peaked earlier than the southern hemisphere. Overall, observational data series (of last 110 years) of sunspot area shows a time gap of 6 to 24 months between the occurrence timing of peak activity in the northern and the southern hemispheres.

\begin{figure}[!h]
\centering
\includegraphics[height=5.5cm, width=8.0cm]{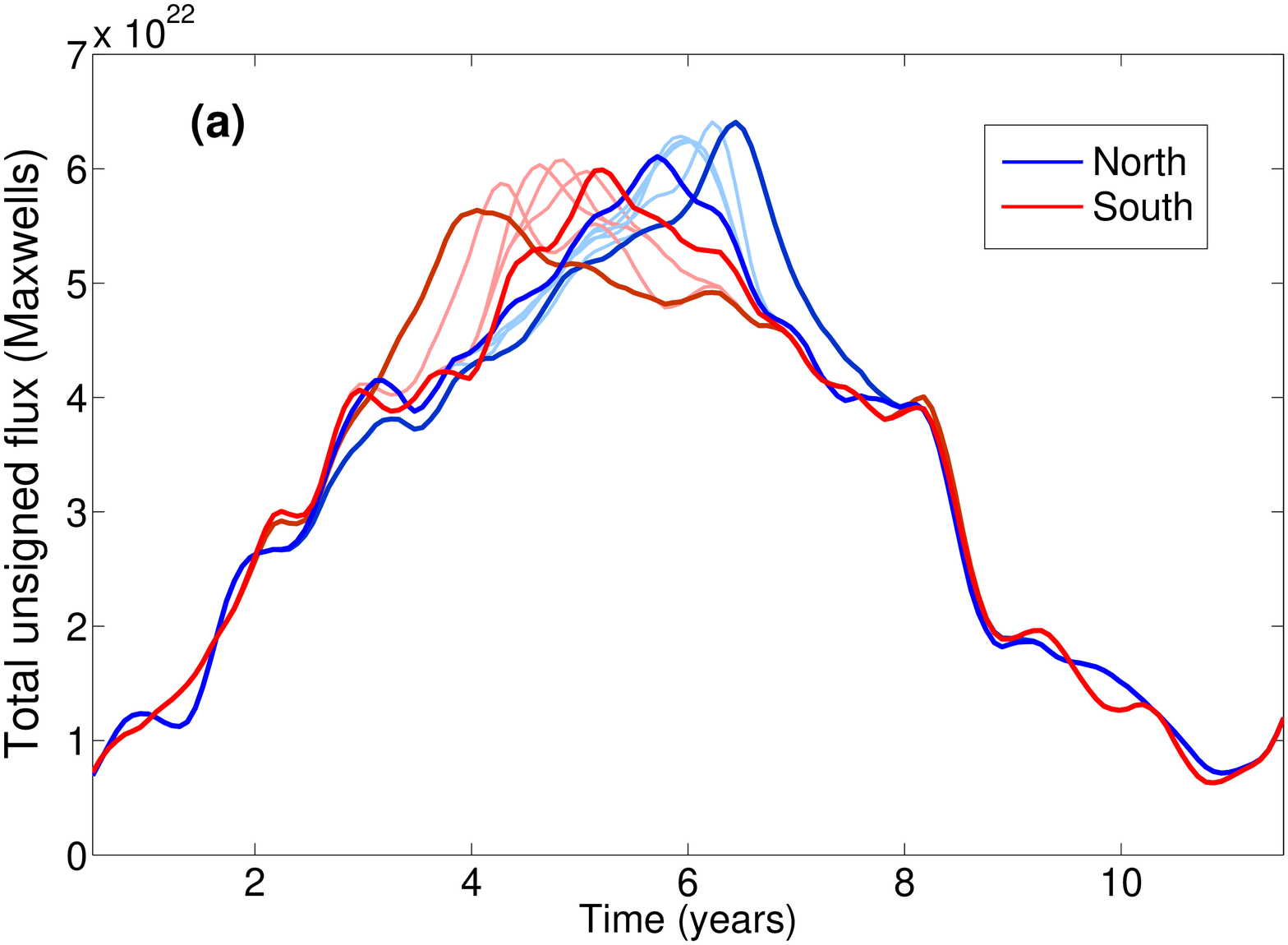}
\includegraphics[height=5.5cm, width=8.0cm]{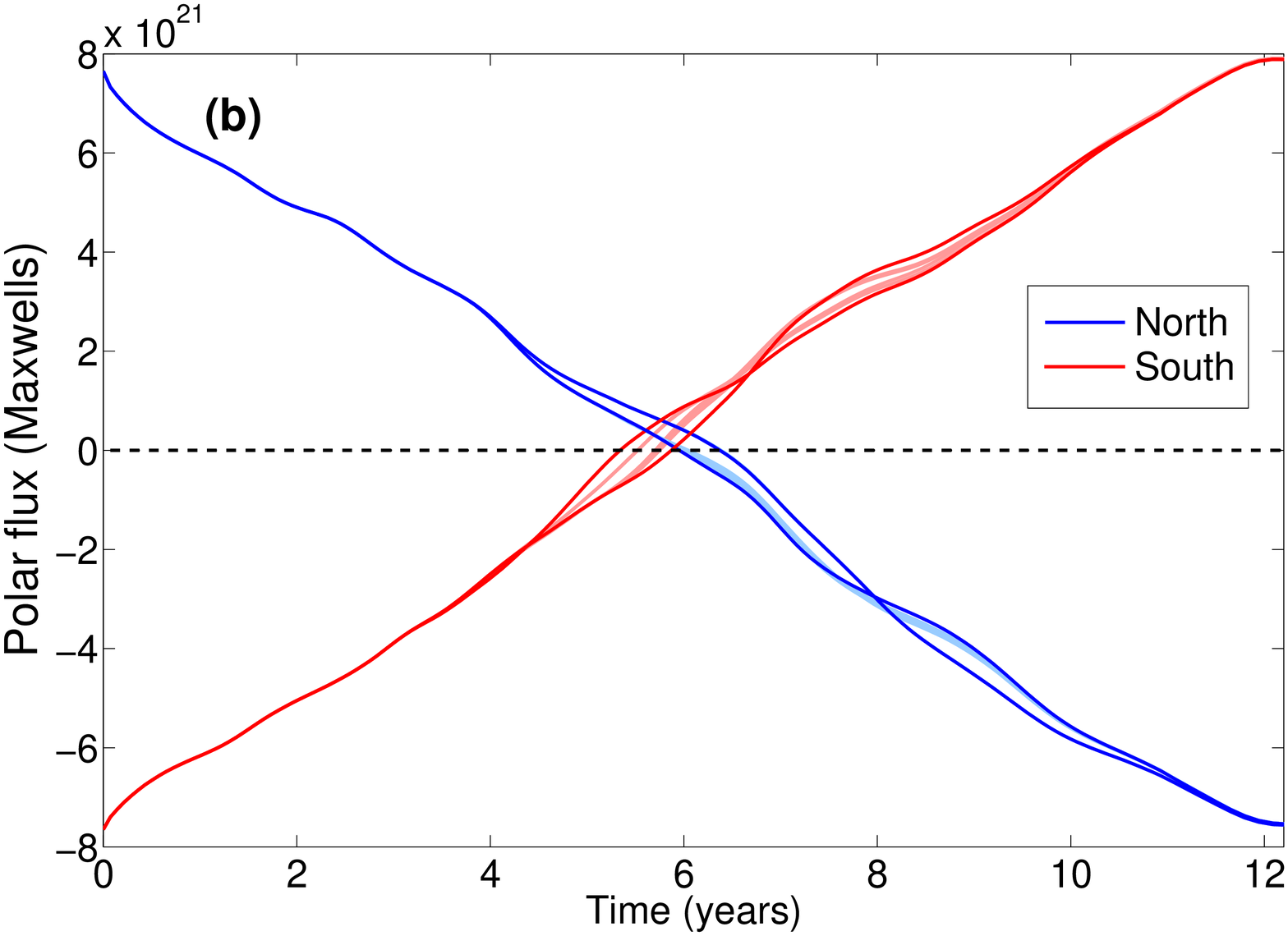}
\includegraphics[height=5.5cm, width=8.0cm]{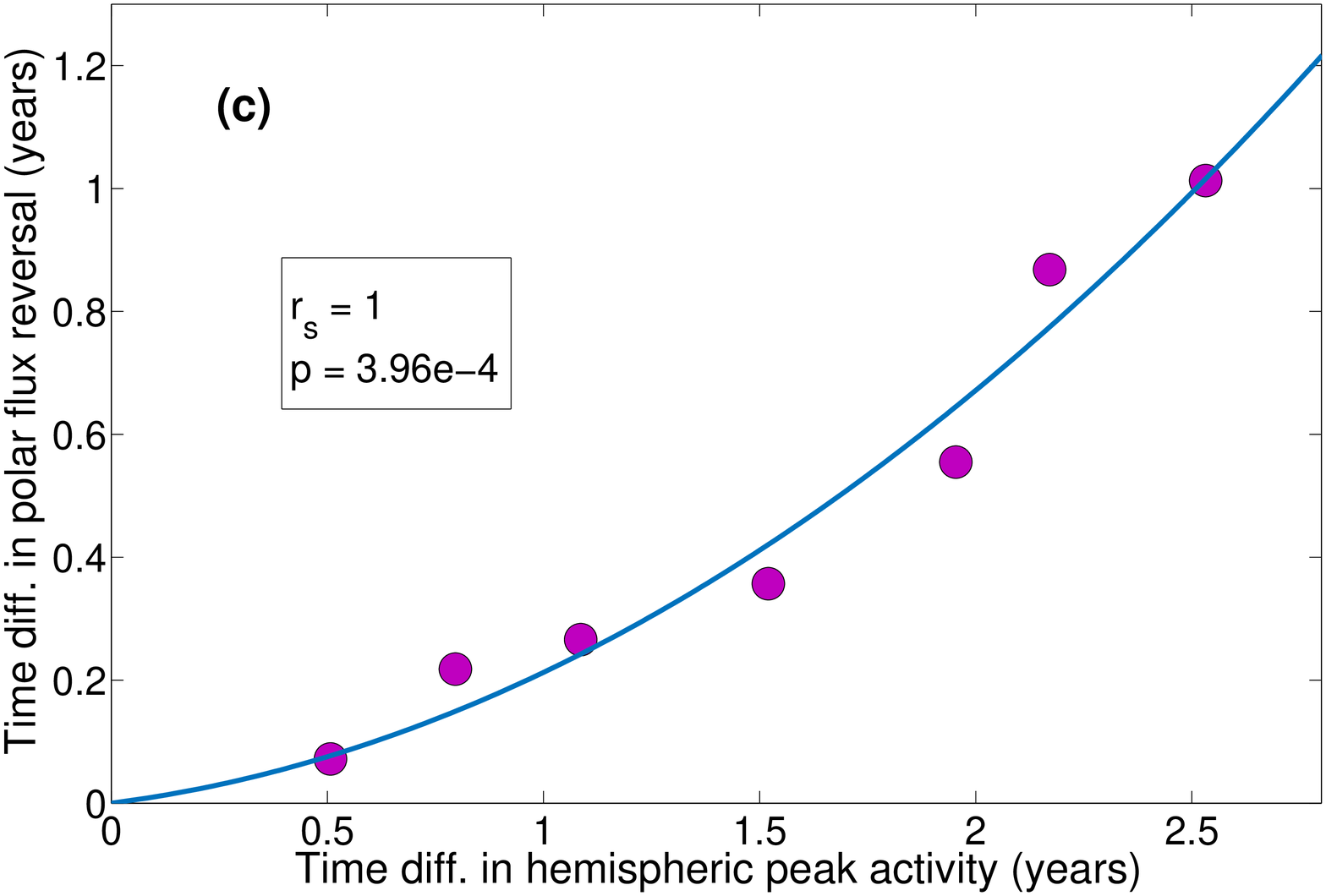}
\caption{(a) represents time evolution of the unsigned magnetic flux associated with the synthetic sunspot input profiles in the northern and southern hemispheres with time difference in their respective peak activity. In figure (b), the evolution of corresponding polar flux in two hemispheres is depicted. (c) The time lag in peak hemispheric activity is compared with the corresponding time difference in the reversal of polar flux, where the blue curve depicts a polynomial (of degree 2) fit to the data points. The associated Spearman's rank correlation coefficient is given with p-value.} 
\label{tim_asym}
\end{figure}

We explore the effect of hemispheric asymmetry on the reversal timing of polar flux as well as its final amplitude at solar minimum by utilizing multiple synthetic sunspot profiles with different time lags. During a specific cycle, the total amount of unsigned magnetic flux associated with the set of emerging sunspots is equal in both the hemispheres. The only asymmetry is introduced through a phase lag between the epochs of peak activity in two hemispheres which varies over a range of 6 to 30 months. We note that the phase dependent mean latitudinal positions of sunspots are roughly similar in two hemispheres such that we can ignore the hemispheric asymmetry regarding the emergence latitudes of sunspots. In Fig.
\ref{tim_asym}(a), time evolution of unsigned flux ($\Phi^{N/S}$) associated with the multiple sunspot profiles are depicted separately with different colored curves for the northern (blue) and the southern hemispheres (red). The corresponding time evolution of hemispheric polar flux [see, Fig. \ref{tim_asym}(b)] demonstrates a spread in the reversal timing of hemispheric polar flux, however, leaving the final amplitude at cycle minimum unaffected. We note that all other model parameters used in the SFT simulations were kept fixed throughout these analyses.

We find a strong correlation between the phase lag in peak of sunspot flux and the corresponding time difference of polar flux reversal, such that Pearson correlation coefficient ($r_p$) is 0.9507 (with a p-value 4.46e-4), and Spearman rank correlation coefficient ($r_s$) is 1 (with a p-value 3.96e-4). Since the generation of polar field through advection and dissipation of surface magnetic field is a complex process, the indication of nonlinearity (i.e., $r_s > r_p$) is entirely expected. We find the relation to be quadratic as depicted by a polynomial (of degree 2) fit of the data set [see, Fig. \ref{tim_asym}(c)]. Although the SFT model used in our study a linear model, we anticipate that the nonlinear behavior is caused by the simultaneous participation of multiple active regions in the flux dissipation process on the solar surface.

\subsection{Hemispheric Asymmetry in Total Sunspot Flux}
\begin{figure}[!h]
\centering
\includegraphics[height=4.8cm, width=8.0cm]{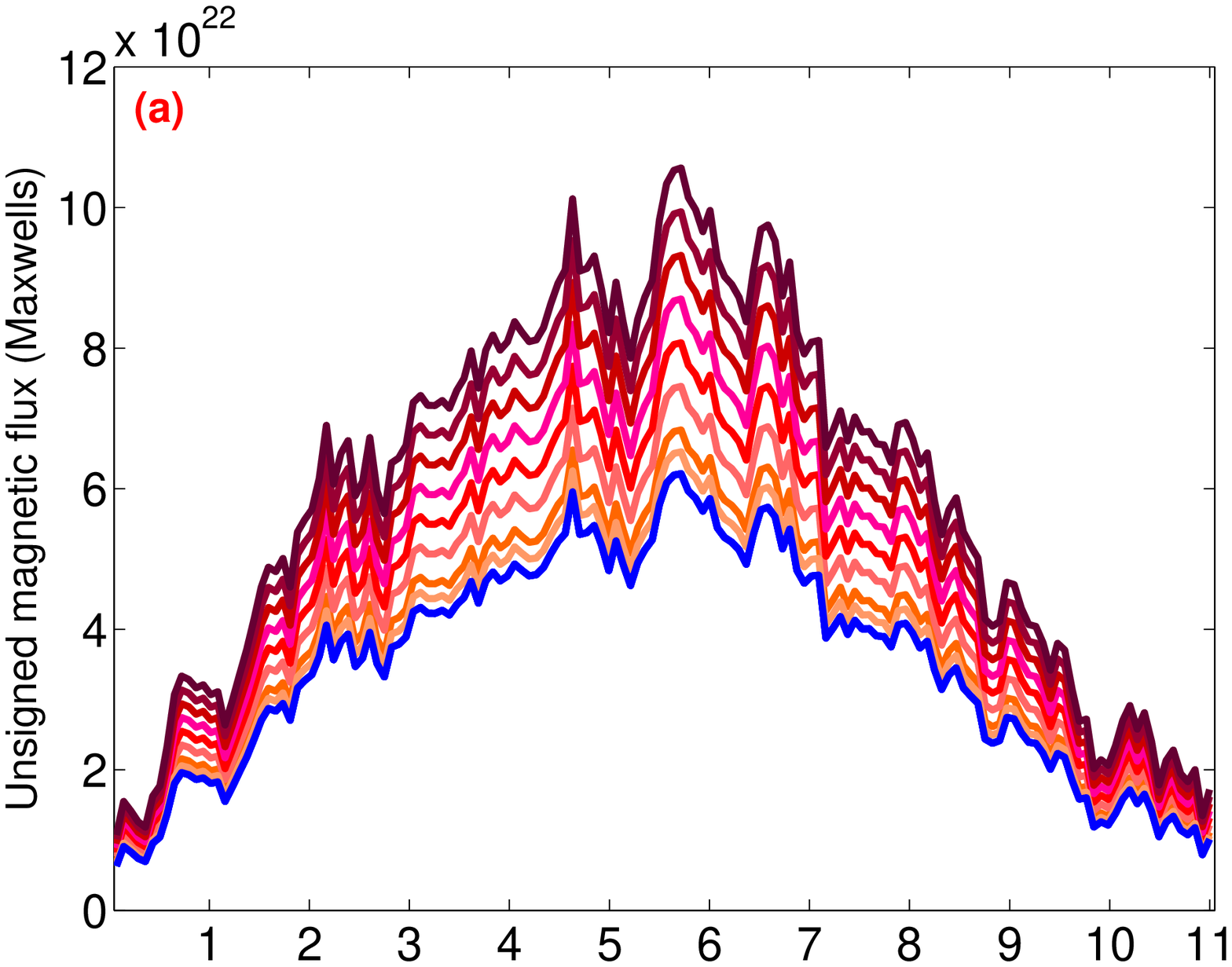}
\includegraphics[height=4.8cm, width=8.0cm]{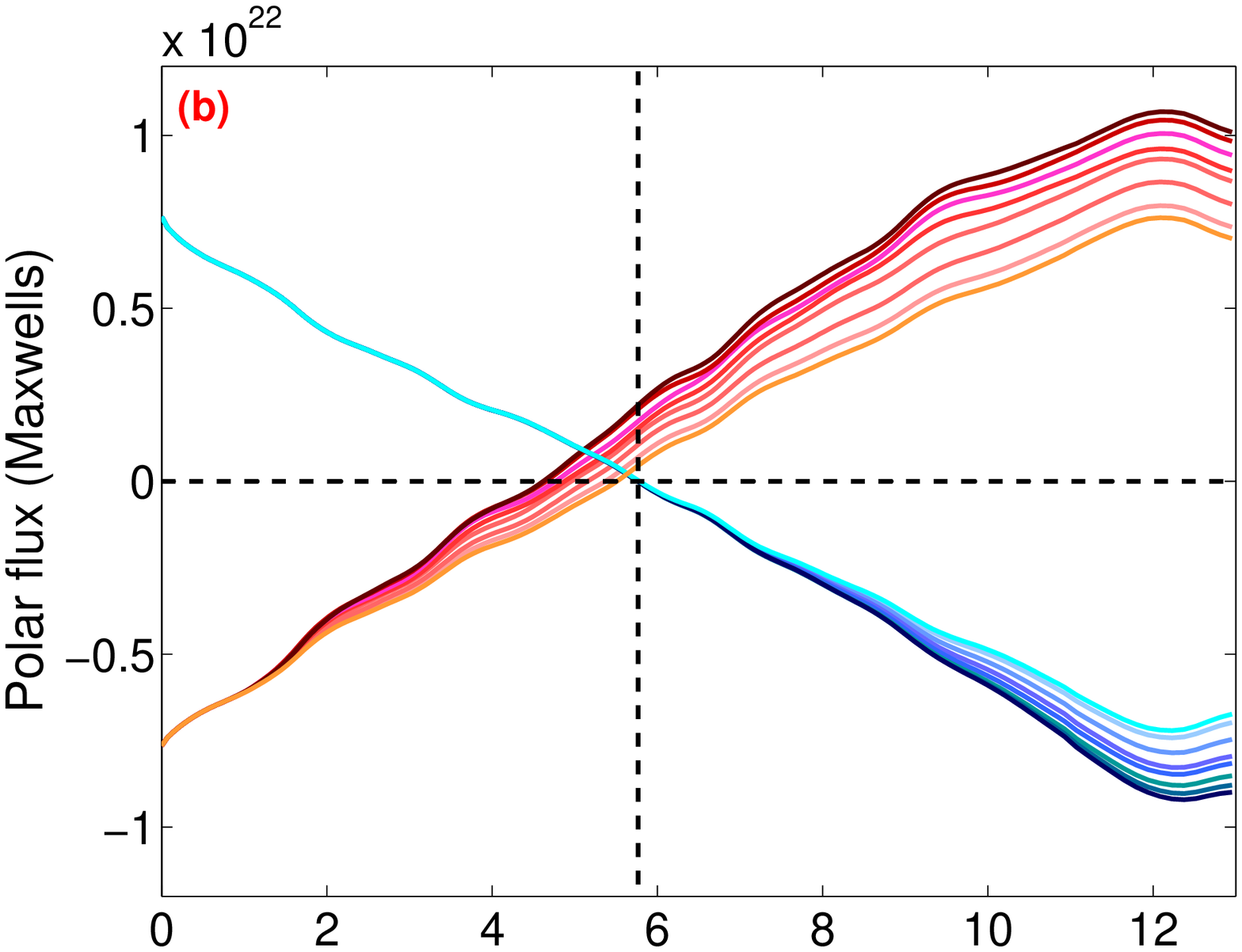}
\includegraphics[height=5.1cm, width=8.0cm]{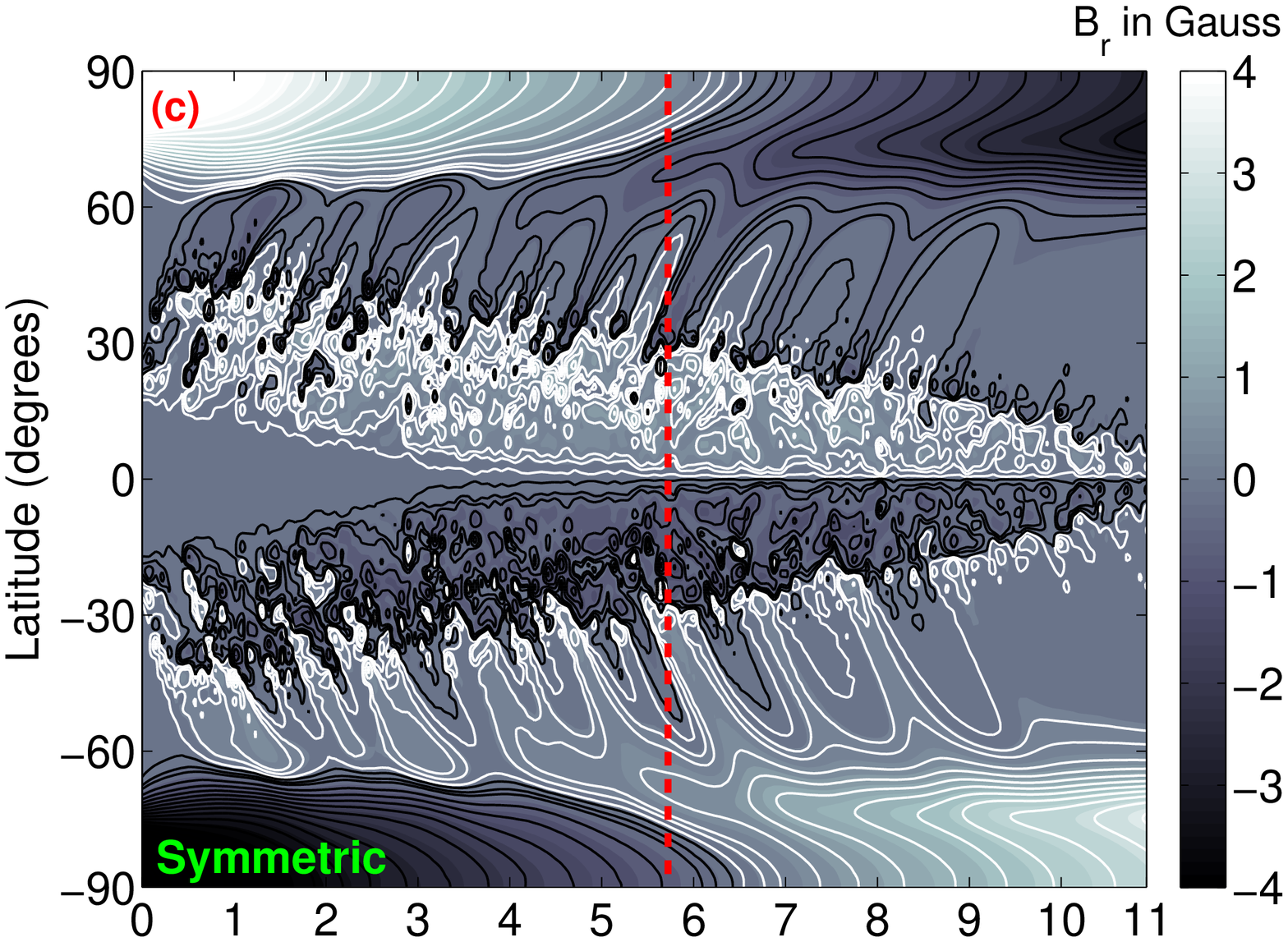}
\includegraphics[height=5.1cm, width=8.0cm]{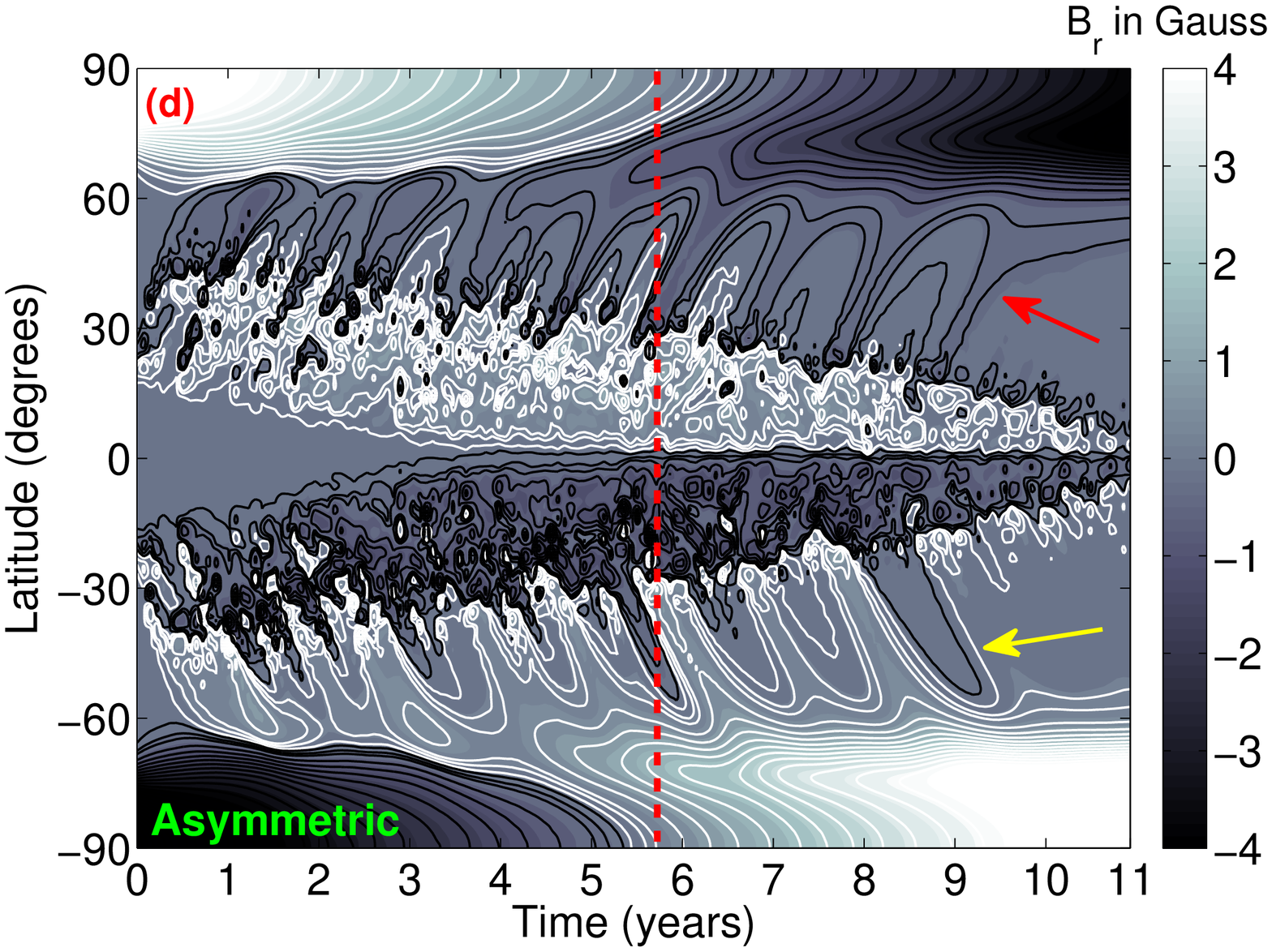}
\caption{(a) Represents the time evolution of unsigned magnetic flux in the northern by a blue curve and in the southern hemispheres by red curves with different shades corresponding to increasing strength (5$\%$ to 70$\%$ compared to the norther hemisphere). (b) Shows the corresponding polar flux evolution in two hemispheres. The time evolution of the longitudinally averaged magnetic field (radial component) as a function of latitude is depicted for two cases: (c) the input sunspot profile is identical in two hemispheres and (d) the southern hemispheric sunspot flux is 70$\%$ higher compared to that in the northern hemisphere. The red and the yellow arrows indicate two prominent distinct patterns in the time latitude distribution of the magnetic field present corresponding to the hemispheric asymmetric case. The red [and the black in (b)] dashed vertical lines refer to the timing of sunspot cycle maximum.} 
\label{flux_asym}
\end{figure}

In the second scenario of our study, the emergence timing and position of individual sunspot appearing in the northern and southern hemispheres are identical to each other. The only difference is that the southern hemispheric spots are larger in areal coverage (thus, have higher magnetic flux) compared to the northern hemispheric spots -- resulting in a surplus of total flux in the southern hemisphere. For simplicity of analysis, we multiplied the area (and also the flux) of each active region in the southern hemisphere by a constant factor `$m$', where $m$ varies over a range of 1.05--1.70, such that the total, as well as the peak amplitude of southern hemispheric flux, becomes 5--70$\%$ more compared to the northern hemispheric flux [see, Fig. \ref{flux_asym}(a)]. Although 70$\%$ asymmetry may seem unusually high given that observational studies \citep{2014SSRv..186..251N} suggest the maximum asymmetry regarding the peak hemispheric activity to be about 20$\%$, we find that in case of solar cycle 20, the total sunspot associated area (and also the flux) in the northern hemisphere was 45$\%$ more compared to that in the southern hemisphere. Additionally we consider a certain reduction in the mean tilt angle of spots associated with the hemisphere with higher activity, although the latitudinal positions of sunspots in two hemispheres are the mirror image of each other. Such modification of mean tilt angle based on cycle amplitude has been observationally reported \citep{2010A&A...518A...7D} and utilized in data-driven SFT simulations \citep{2010ApJ...719..264C,2011A&A...528A..82J,2014ApJ...791....5J}. 

Based on analysis of observational data of sunspot tilt angles, \cite{2011A&A...528A..82J} describes how a factor [T$_n$, see, equation (15) in their paper] dependent on the cycle amplitude (in terms of sunspot number) can be introduced to calculate tilt angle of BMRs. Since peak sunspot number and sunspot flux (or area) are well correlated, we utilize this concept to establish a similar relation between the average tilt angle and peak amplitude of unsigned hemispheric magnetic flux (monthly averaged), $T^{N/S} = 1.71-0.12mF^{N/S}$. Here, $F^{N/S}$ represents the peak amplitude of unsigned magnetic flux ($\Phi^{N/S}$) associated with sunspots appearing in the northern and the southern hemispheres, respectively. The factor $T^{N/S}$ is incorporated in determining the tilt angle of an individual active region through the relation $\alpha = g T^{N/S} \sqrt{|\lambda|}$, where $\alpha$, $g$ and $\lambda$ represent the same quantities as described in section 3.1.3. 

The corresponding time evolution of polar flux shows that an increase in sunspot associated magnetic flux in the southern hemisphere results in an early reversal and an increment in the final strength (at cycle minimum) of the southern hemispheric polar flux [see, the set of red curves in Fig. \ref{flux_asym}(b)]. Additionally, the final polar flux in the northern hemisphere also increases [see, the set of blue curves in Fig. \ref{flux_asym}(b)], although, the sunspot associated flux does not change in the northern hemisphere while considering different synthetic sunspot profiles. In Fig. \ref{flux_asym}(c), we represent the time-latitude distribution of a longitudinally averaged radial component of the surface magnetic field (also known as magnetic butterfly diagram) corresponding to the symmetric case where the emergence profile of sunspots are identical in two hemispheres. In Fig. \ref{flux_asym}(d), we depict the magnetic butterfly diagram corresponding to an input profile with the southern hemisphere being 70$\%$ more active compared to the north. With increased input sunspot flux, a proportionate increment in the southern hemispheric polar flux is quite expected, but the cause of enhancement in the northern hemispheric polar flux lies in the complexity of the B-L mechanism. We speculate that the high magnetic flux content of leading polarity spots in the southern hemisphere facilitate higher cross-equatorial flux cancellation with leading polarities belonging to the northern hemisphere. This eventually reduce the scope of intra-hemispheric flux cancellation among leading and following polarities of BMRs in the northern hemisphere and increases the amount of resultant unipolar flux that is accumulated from the following polarity spots and subsequently is advected towards the north pole. The magnetic field distribution of negative polarity in the northern hemisphere [marked by black contour lines in Fig. \ref{flux_asym}(c) and (d)] shows a significant contrast between the symmetric and asymmetric cases and indicates larger transportation of magnetic flux (of negative polarity) towards the pole in the asymmetric case [for example, see the structure pointed by the red arrow in Fig. \ref{flux_asym}(d)]. We note that the effect of cross-equatorial flux cancellation among leading spots (of opposite magnetic polarity) becomes more pronounced after cycle maximum (marked by the red dashed line) has occurred.

Interestingly, an imbalance in activities between two hemispheres can result in advection of magnetic flux from both the leading and following polarities towards the polar region -- effectively reducing the net polar flux in the dominant hemisphere. A manifestation of this phenomenon can be observed in the southern hemisphere where negative polarity flux from the leading polarity spots reach beyond $-$55$^{\circ}$ latitude [pointed by the yellow arrow in Fig. \ref{flux_asym}(d)]. 

As polar field is generated through a complex process of flux cancellation and advection, we find the nature of positive correlation that exists between the cycle amplitude and the final strength of the polar flux at cycle minimum to be non-linear in the southern hemisphere. The corresponding Spearman rank correlation coefficient is 1.0 with a p-value of 99.99 $\%$. A quantitative analysis shows that a 70$\%$ increment of magnetic flux associated with the sunspots in the southern hemisphere (compared to the north) results in only 14$\%$ hemispheric asymmetry in the final amplitude of polar flux during cycle minimum. 

\subsection{Scatter in Active Region Tilt Angle}
Another factor capable of introducing hemispheric irregularities is the randomness present in the tilt angle distribution of active regions emerging on the photosphere and participating in the B-L mechanism. Several observational studies \citep{2010A&A...518A...7D,1999SoPh..189...69S,2013SoPh..287..215M} have found a significant scatter in tilt angles of BMRs, in addition to the deterministic latitude-dependent tilt angles. Within the solar convection zone, the Coriolis force acts on a diverging flow field in a buoyantly rising flux tube and results in a systematic latitudinal tilt of active region as delineated by Joy's law \citep{1993A&A...272..621D,1993ApJ...405..390F,1995ApJ...438..463F,2009LRSP....6....4F,2013SoPh..287..239W}. The turbulent convective flows on the rising flux tube introduces a randomness which is inversely proportional to its magnetic field strength \citep{1995ApJ...438..463F,1996ApJ...458..380L,2011ApJ...741...11W,2013SoPh..287..239W}. In this section, we explore the impact of scatter present in tilt angle of active regions on the final hemispheric polar flux amplitude and the associated asymmetry during cycle minimum.

A study utilizing SFT simulations \citep{2004A&A...426.1075B} has shown the final polar flux to be proportional to the tilt angle of active regions. \cite{2014ApJ...791....5J} and \cite{2017SoPh..292..167N} have demonstrated that large sunspots with large scatter in their tilt angles can significantly affect polar field amplitude during solar minimum. Moreover, a large individual sunspot of non-Hale nature appearing at lower latitude can potentially reduce the polar field strength \citep{2015SoPh..290.3189Y}. To find the possible scatter in tilt angle of a particular active region, we follow the prescription given by \cite{2014ApJ...791....5J}, where they have established an empirical (linear logarithmic) relation between the variance of tilt angle distribution and the associated active region area [see, equation (1) in their paper] by analyzing observational data. We consider a sunspot input profile with the emergence timing and position on the solar surface to be identical in the northern and the southern hemispheres; the only difference is introduced through the randomness of tilt angles. The tilt angle of every single active region is determined by the relation, $\alpha = g \sqrt{|{\lambda}|}+\epsilon$, where the latitude-dependent Joy's law decides the first part in the right-hand side and the second part, $\epsilon$, represents the randomness \citep{2014ApJ...791....5J}. The value of $\epsilon$ is chosen through random selection from a Gaussian distribution with zero mean and a standard deviation decided by the area of the active region in consideration. Tilt angle of every active region in the symmetric input profile is modified by adding individual $\epsilon$ selected through the above process -- eventually generating a sunspot input profile with hemispheric asymmetry. We study the evolution of polar flux in both the hemispheres while considering 50 such distinct input profiles.

\begin{figure}[!h]
\centering
\includegraphics[height=5.4cm, width=9.0cm]{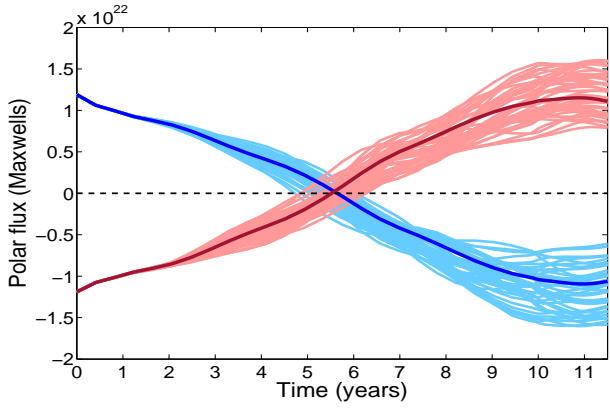}
\caption{Time evolution of polar flux in the northern and the southern hemispheres associated with the 50 individual input profiles are depicted by the set of light blue and light red curves, respectively, whereas the dark blue and dark red curves represent the same corresponding to the standard symmetric profile.} 
\label{tilt}
\end{figure}

Scatter in active regions' tilt angle results in significant uncertainties in the final polar field strength during solar minimum in both the hemispheres (see, Fig. \ref{tilt}). While the timing of polarity reversal and the final amplitude of the hemispheric polar flux are identical for the symmetric profile [the dark blue and dark red curves in Fig. \ref{tilt}], we see a spread of $\pm$45$\%$ and $\pm$35$\%$ in the northern and the southern hemispheric polar flux at cycle minimum, respectively, corresponding to the 50 asymmetric sunspot input profiles. Furthermore, the randomness in tilt angle affects the reversal timing of the polar flux -- resulting in uncertainty of $\pm$ 8 months in both the hemispheres (on average) with respect to the timing related to the standard symmetric profile. Among these 50 realizations, hemispheric asymmetry in the final polar flux strength becomes as high as 4.2$\times$10$^{21}$ Maxwells which is 36$\%$ of the polar flux amplitude obtained by using the standard symmetric profile. 

\subsection{Hemispheric Asymmetry in Transport Parameters}
Transport parameters involved in the B-L mechanism are magnetic diffusivity and two large-scale velocities -- meridional circulation and differential rotation. Among these three parameters, hemispheric asymmetry can exist only in the velocity fields, as diffusion originated from turbulent motion of the convective cells within the solar convection zone are homogeneous and isotropic \citep{2006JGRA..11112S01N}. Observational studies (\citealt{2013A&A...552A..84Z} and the references therein) have found the rotation rates to vary around the mean profile [see, equation (3)] about 3--4$\%$ in two hemispheres, such that the maximum hemispheric asymmetry that can exist in differential rotation is about 8$\%$. However, inclusion of such small variation of differential rotation in the SFT simulations does not induce any significant hemispheric asymmetry (less than 1$\%$) in the reversal timing and the final amplitude of polar field (during cycle minimum). 

\begin{figure}[!h]
\centering
\includegraphics[height=4.9cm, width=8.0cm]{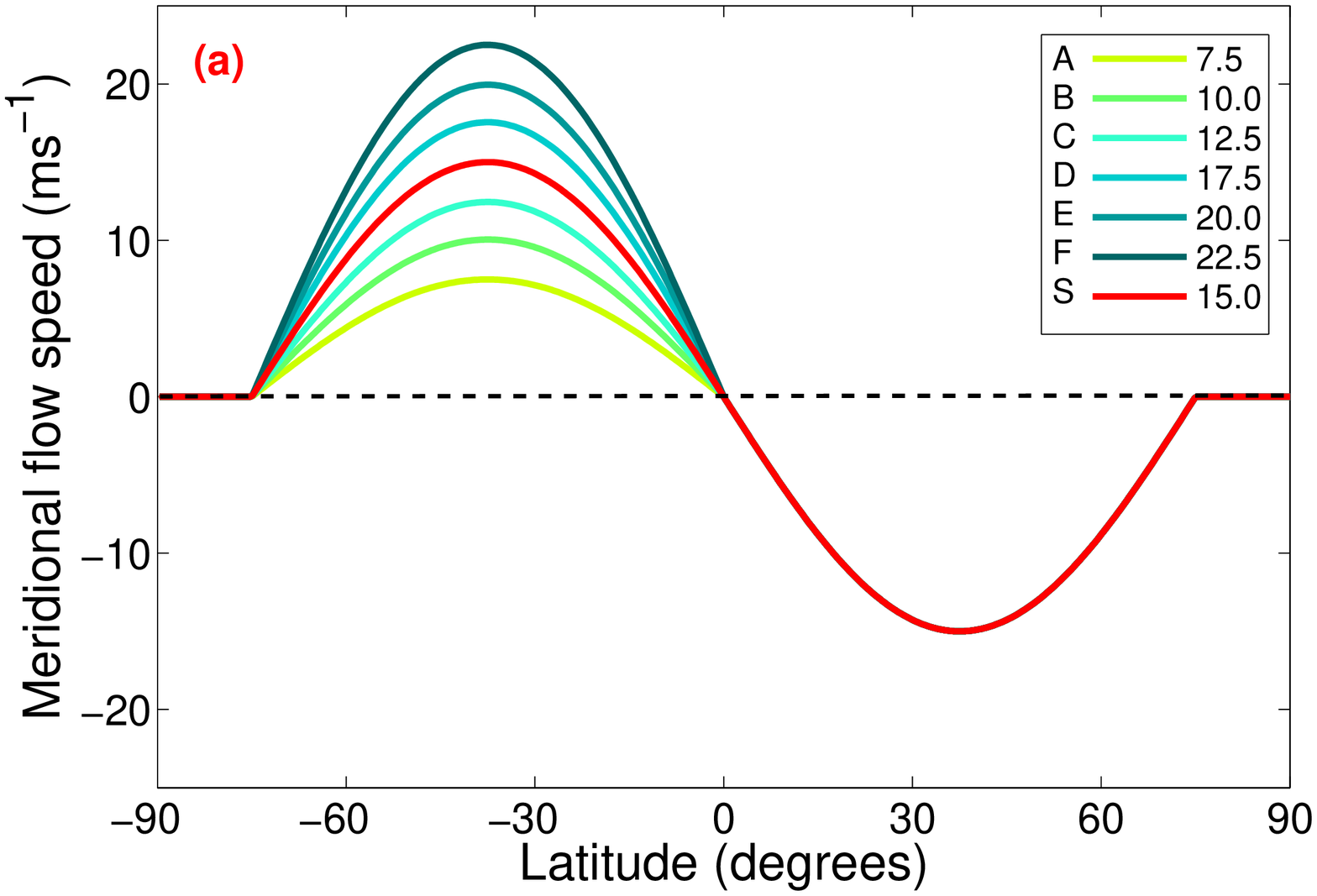}
\includegraphics[height=5.0cm, width=8.0cm]{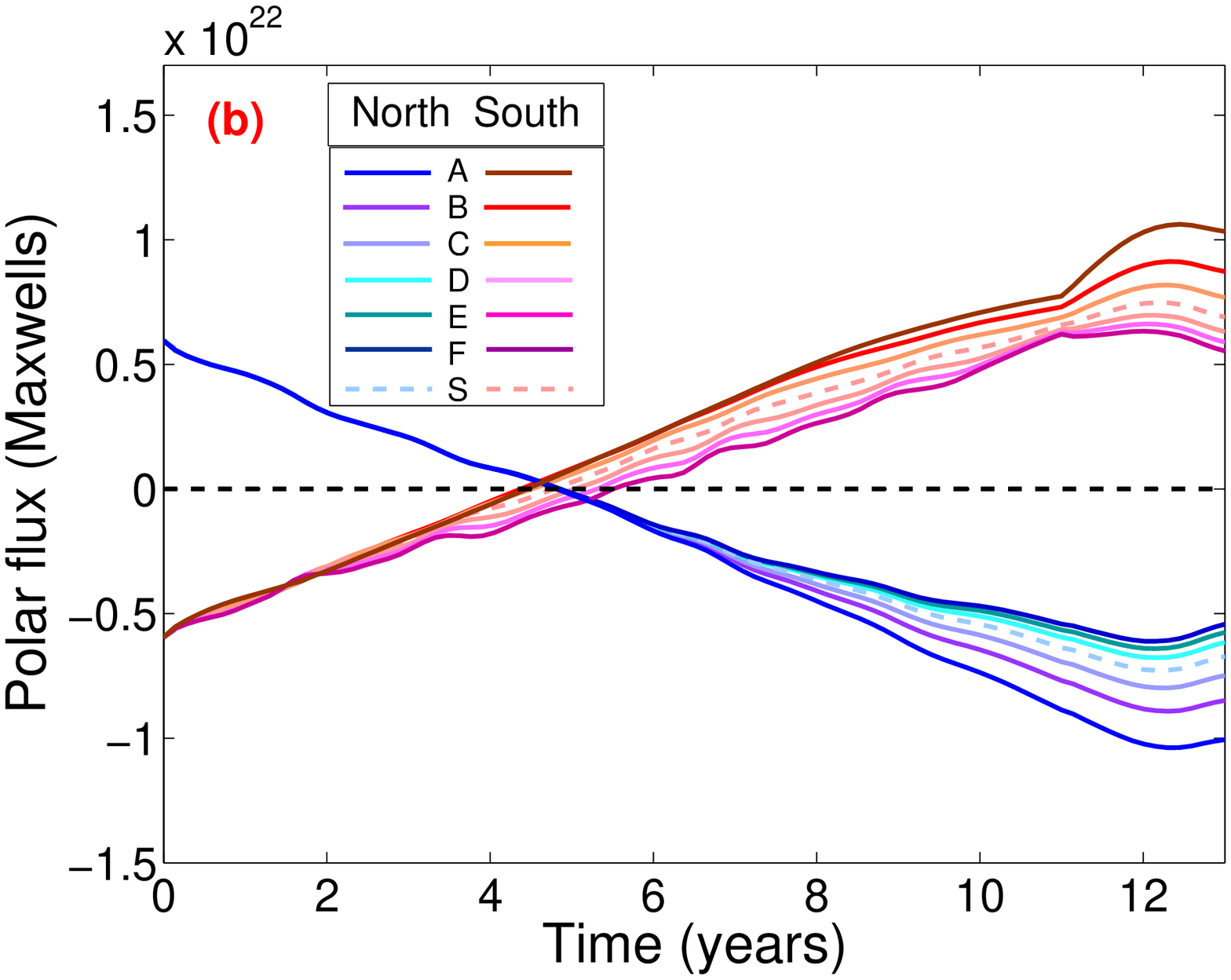}
\includegraphics[height=5.1cm, width=8.0cm]{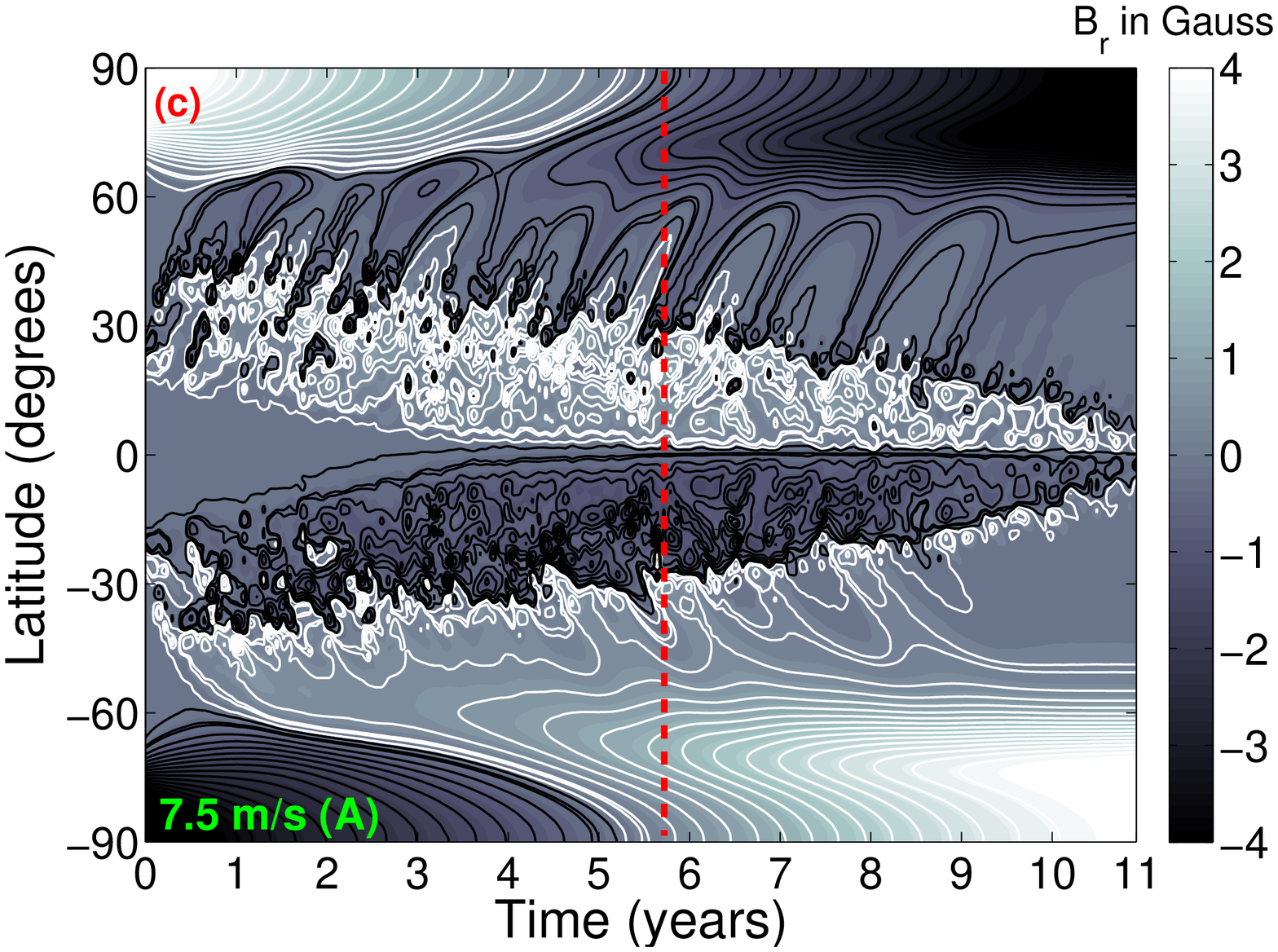}
\includegraphics[height=5.1cm, width=8.0cm]{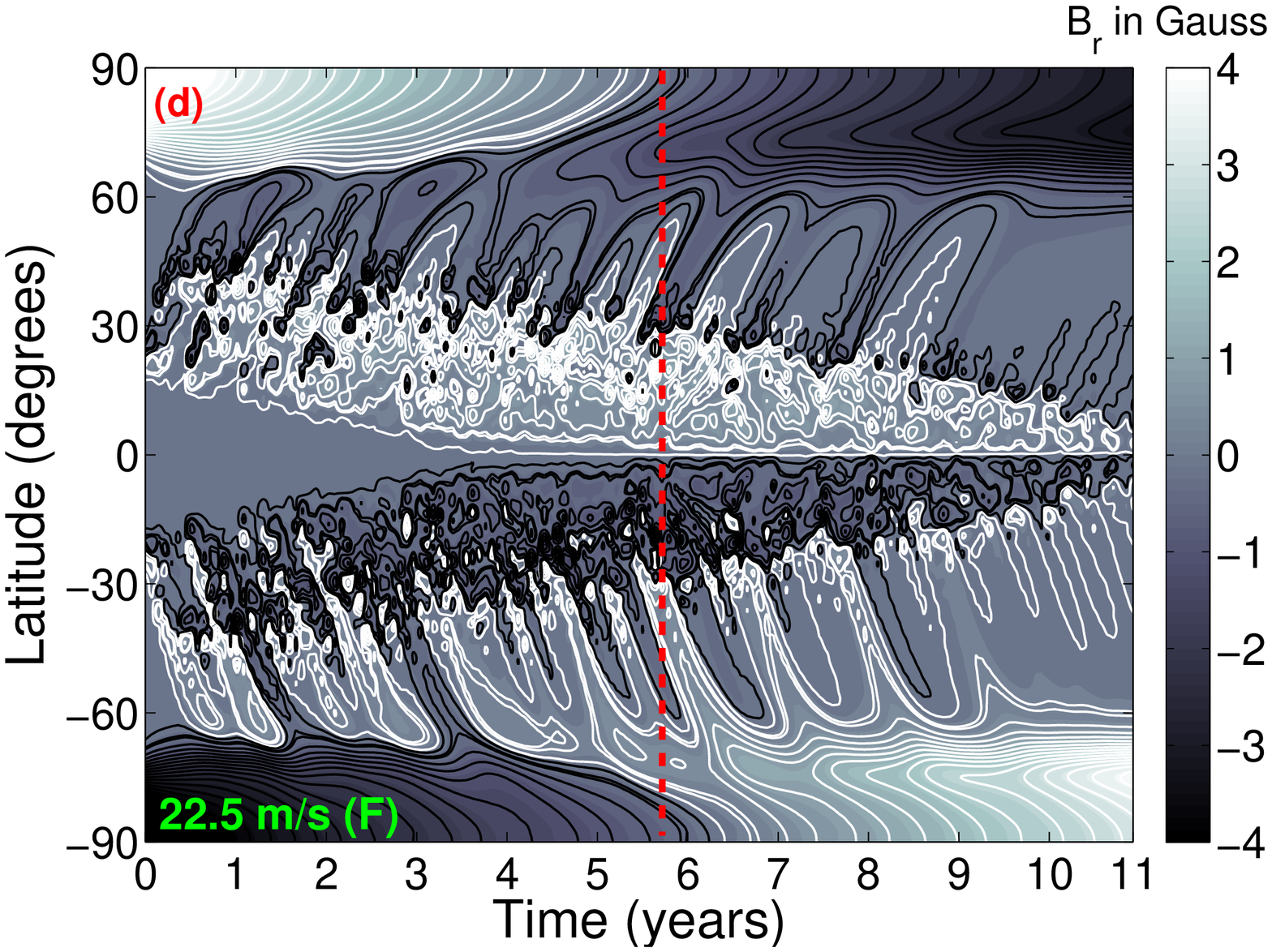}
\caption{(a) Different meridional circulation profiles are depicted as functions of latitude where the peak speed varies from 7.5 ms$^{-1}$ to 22.5 ms$^{-1}$ (denoted by different colors) in the southern hemisphere. The positive (and the negative) velocities indicates the flow is towards the south (and the north) pole. Each flow profile is labeled alphabetically staring from `A' to `F', while `S' represents the symmetric profile. (b) Depicts the time evolution of the corresponding hemispheric polar flux. Sub-figures (c) and (d) represent the magnetic butterfly diagram corresponding to the case A (with peak flow speed 7.5 ms$^{-1}$) and the case F (peak speed 22.5 ms$^{-1}$). The red dashed vertical lines depict the time of sunspot maximum.} 
\label{flow_asym}
\end{figure}

Helioseismology \citep{2002ApJ...570..855H,2006SoPh..236..227Z,2015SoPh..290.3113K} as well as feature tracking techniques \citep{2010Sci...327.1350H,2011ApJ...729...80H} have revealed that the observationally deduced meridional circulation profile on the solar surface occasionally differs from its time-invariant and simplistic antisymmetric (about the equator) form [as expressed by equation (4)] and exhibits variation over time along with hemispheric asymmetry. In this study, we primarily explore the consequences of different peak flow speeds in two hemispheres on the evolution of polar field without introducing any time-varying component in the meridional circulation profile. We vary the peak speed of meridional flow in the southern hemisphere [i.e., the amplitude of $v_0$ in equation (4)] within a range of 7.5 -- 22.5 ms$^{-1}$ while keeping the peak flow speed fixed at 15 ms$^{-1}$ in the northern hemisphere -- effectively introducing a $\pm$50$\%$ north-south asymmetry in the amplitude of the flow [see, Fig. \ref{flow_asym}(a)]. 

We consider six different meridional flow profiles [case A to F, see, Fig. \ref{flow_asym}] and study the time evolution of polar flux by running SFT simulations with a hemispherically symmetric sunspot input profile. From Fig. \ref{flow_asym}(b), we observe that with decreasing flow speed in the southern hemisphere, polarity reversal occurs earlier and also results in stronger final polar flux during cycle minimum in the same hemisphere. Surprisingly, the change in peak flow speed in the southern hemisphere profoundly affects the final amplitude of the northern hemispheric polar flux. In order to explain these features we analyze the magnetic butterfly diagram associated with two extreme cases, A and F, where the peak flow speed in the southern hemisphere is 50$\%$ lesser (i.e., 7.5 ms$^{-1}$) and 50$\%$ higher (22.5 ms$^{-1}$), respectively, compared to the peak flow speed in the northern hemisphere [see, Fig. \ref{flow_asym}(c) and \ref{flow_asym}(d)]. 

Evolution of the surface magnetic field is governed by the interplay between velocity fields and magnetic diffusivity. While the primary role played by meridional circulation is to carry magnetic flux from the equator to the polar region, magnetic diffusion promotes cancellation of magnetic flux among opposite polarities along with participating in the process of advection of flux. During the initial phase of a sunspot cycle, active regions primarily appear at the higher latitudes in both the hemispheres and intra-hemispheric interaction occurs between the leading and the following spots. Thus, magnetic flux cancellation remains restricted within individual hemispheres during this phase. As sunspot activity belts in two hemispheres approach towards the equator (predominantly visible after sunspot maximum), cross-equatorial flux cancellation among the leading spots belonging to the northern and the southern hemispheres becomes important. Throughout the cycle, the residual flux from the following polarity spots is transported to the poles aided by meridional circulation. 

Therefore, a slower flow in the southern hemisphere (case A) provides sufficient time for a substantial flux cancellation of opposite magnetic polarities -- effectively allowing more unipolar flux to travel towards south pole as observed in Fig. \ref{flow_asym}(c). This results in an early reversal as well as a higher final strength of the southern hemispheric polar flux. The same mechanism of flux cancellation across the equator also instigates a change in the norther hemispheric polar flux. Slower southern hemispheric meridional circulation facilitates a higher amount of unipolar magnetic flux (from the following spots) in the northern hemisphere to advect towards the north pole -- eventually resulting in higher amplitude. On contrary, a faster meridional flow in the southern hemisphere (case F) drags magnetic flux from both the leading and the following spots in such a way that both polarities can reach near the polar region -- effectively delaying the process of polarity reversal and building up of new polar field in that hemisphere. In Fig. \ref{flow_asym}(d), we observe such events to occur in the southern hemisphere during the 2$^{nd}$ and the 4$^{th}$ year, where flux from the leading spots reach beyond $-$60$^{\circ}$ latitudes. The same fast flow hinders cross-equatorial flux cancellation -- eventually reducing the final strength of the north hemispheric polar flux. We note that the distinction between the magnetic field distribution in the northern hemisphere corresponding to cases A and F enhances after cycle maximum is reached, beyond which cross-equatorial flux cancellation becomes profound. In all cases (A to F), the southern hemispheric polar flux is slightly stronger compared to the northern hemisphere. In summary, a hemispheric asymmetry as high as $\pm$50$\%$ in the peak flow speed eventually introduces only about 3$\%$ asymmetry in the final polar field strength associated with two hemispheres. 

So far we have explored different aspects of the B-L mechanism responsible for north-south asymmetry in the hemispheric polar field evolution and investigate their individual potential. We summarize our findings in Table 1.  Our next aim is to study how the asymmetry present in polar flux during cycle minimum is translated and reflected in hemispheric activity of the following cycle in the context of dynamo mechanism.

\section{Translation of Hemispheric Asymmetry in the Succeeding Solar Cycle: A Dynamo Perspective}
Several numerical studies using dynamo simulations have investigated the origin of hemispheric asymmetry and identified two sources: stochastic fluctuation and non-linear effects [discussed in detail in recent reviews by \citealt{2014SSRv..186..251N} and \citealt{2015SSRv..196..101B}]. While nonlinearity is embedded in the dynamo equations, stochastic fluctuations can be infused in the source of poloidal field. Randomness in the flows associated with the convective cells in the turbulent solar convection zone can lead to fluctuations in the mean field-$\alpha$ effect which is considered as a potential source for poloidal field generation from the toroidal component (in azimuthal direction). Moreover, recent observational \citep{2013ApJ...767L..25M} as well as numerical works \citep{2002Sci...296.1671N,2004A&A...427.1019C,2008ApJ...673..544Y,2015Sci...347.1333C,2018NatComm...9...5209B} have established the B-L mechanism as the prime candidate for poloidal field generation which acts through a combination of diffusion, cancellation and advection of magnetic field associated with the tilted active regions emerging on the solar surface. Thus, any sudden variation in the transport parameters associated with the B-L mechanism and significant scatter in the active region tilt angle can cause fluctuation in the poloidal field source term. To explore how the hemispheric asymmetry obtained from the SFT simulations is translated in the succeeding cycle, we consider a dynamo model with both the B-L mechanism and mean field $\alpha$-effect as two sources for poloidal field generation (as described in section 3.2), where the irregularities and fluctuations are introduced only through the B-L source. The origin of hemispheric asymmetry using a dynamo model has been explored recently by \cite{2018A&A...618A..89S}.

We follow the same approach utilized by \cite{2018NatComm...9...5209B} (an approach also similar to the method used by \citealt{2007MNRAS.381.1527J}) to include the SFT-generated surface magnetic field during sunspot minimum in the poloidal field source term of the dynamo model. We calculate the vector potential on the solar surface [say, $A^{SFT}(R_{\odot},\theta,t_{min})$] by integrating $B^{SFT}(R_{\odot},\theta,t_{min})$, which is obtained by averaging the radial component of the surface magnetic field over longitude($\phi$). The relation between these two quantities is described as below \citep{2007MNRAS.381.1527J},
\begin{equation}
B^{SFT}_r(R_{\odot},\theta,t_{min})=\dfrac{1}{R_{\odot} \sin \theta} \dfrac{\partial}{\partial \theta} [\sin \theta A^{SFT}(R_{\odot},\theta,t_{min})]
\end{equation}
In case of $A_{SFT}$ being nonzero at any of the poles, few terms in equation (\ref{pol}) will encounter singularity \citep{2007MNRAS.381.1527J}. To ensure that $A_{SFT}$ becomes zero at both the poles, we use the following relations to calculate $A^{SFT}$ on the solar surface at cycle minimum ($t_{min}$) for two hemispheres separately,
\begin{equation}
  A^{SFT}(R_{\odot},\theta,t_{min}) \sin \theta=
 \begin{cases}
    \int_{0}^{\theta} B_r(R_{\odot},\theta^{\prime},t_{min}) \sin \theta^{\prime} d\theta^{\prime}, 0 \leq \theta < \pi / 2\\
   \int_{\pi}^{\theta} B_r(R_{\odot},\theta^{\prime},t_{min}) \sin \theta^{\prime} d\theta^{\prime}, \pi/2 < \theta \leq \pi.
  \end{cases}
\end{equation}

The task of comprising the $A^{SFT}(R_{\odot},\theta,t_{min})$ in the magnetic vector potential (i.e., the poloidal field source, $A^{Dyn}$) associated with the dynamo simulation is conducted through the following process. We first evaluate a function $\gamma (\theta)$ by taking a ratio between $A^{SFT}(R_{\odot},\theta,t_{min})$ and $A^{Dyn}(R_{\odot},\theta,t_{min})$ [as shown in Fig.7(b)]. The imprint of the B-L mechanism as simulated by the SFT model is infused in this $\gamma(\theta)$ function. $\gamma (\theta)$ is comprised of two parts, a constant factor `$c$' and a latitude dependent function $\zeta(\theta)$. The constant, `$c$' arises due to the difference in amplitude of $A^{SFT}$ and $A^{Dyn}$ obtained from two disparate numerical models. The value of `$c$' remains unaltered while incorporating individual $A^{SFT}$'s in dynamo simulation. The other function, $\zeta(\theta)$ takes care of the distinct latitudinal distribution of $A^{SFT}$ and $A^{Dyn}$ on the solar surface. Thus, the latitudinal variation of $\zeta(\theta)$ is subjected to the particular $A^{SFT}$ in consideration. Assuming the B-L mechanism as a near-surface process, we modify $A^{Dyn}$ by multiplying it with $\gamma(\theta)$ within a restricted region spanning from 0.8$R_{\odot}$ to $R_{\odot}$ (over the full range of latitude). Finally, the modified $A^{Dyn}$ is used as an initial condition at cycle minimum to perform the dynamo simulation. We compare the quantity $B^{Dyn}$ (as described in section 3.2) associated with the northern and southern hemispheres to explore hemispheric asymmetry.

\begin{figure}[!h]
\centering
\includegraphics[height=5.0cm, width=7.5cm]{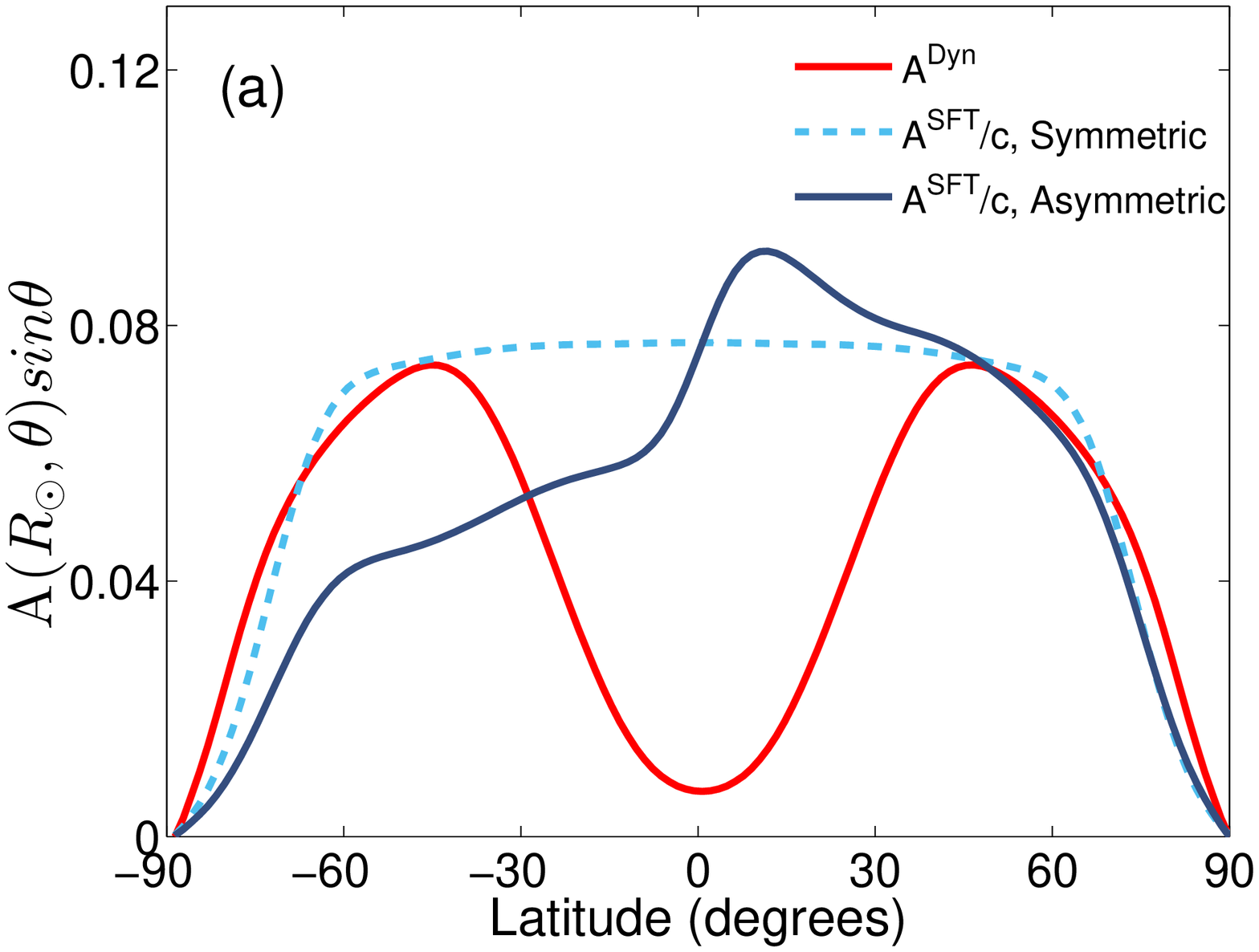}
\includegraphics[height=5.5cm, width=7.0cm]{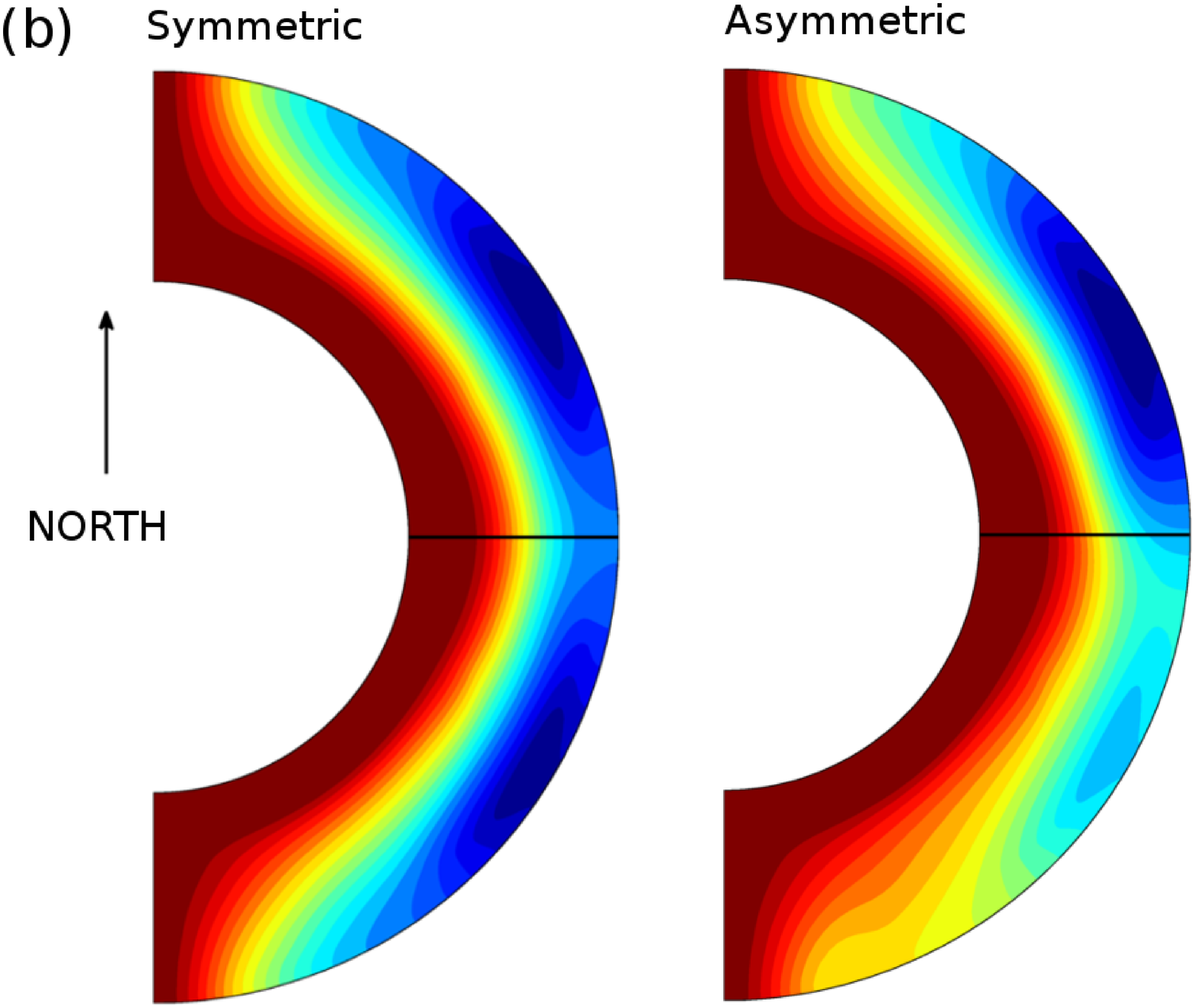}
\includegraphics[height=5.0cm, width=7.5cm]{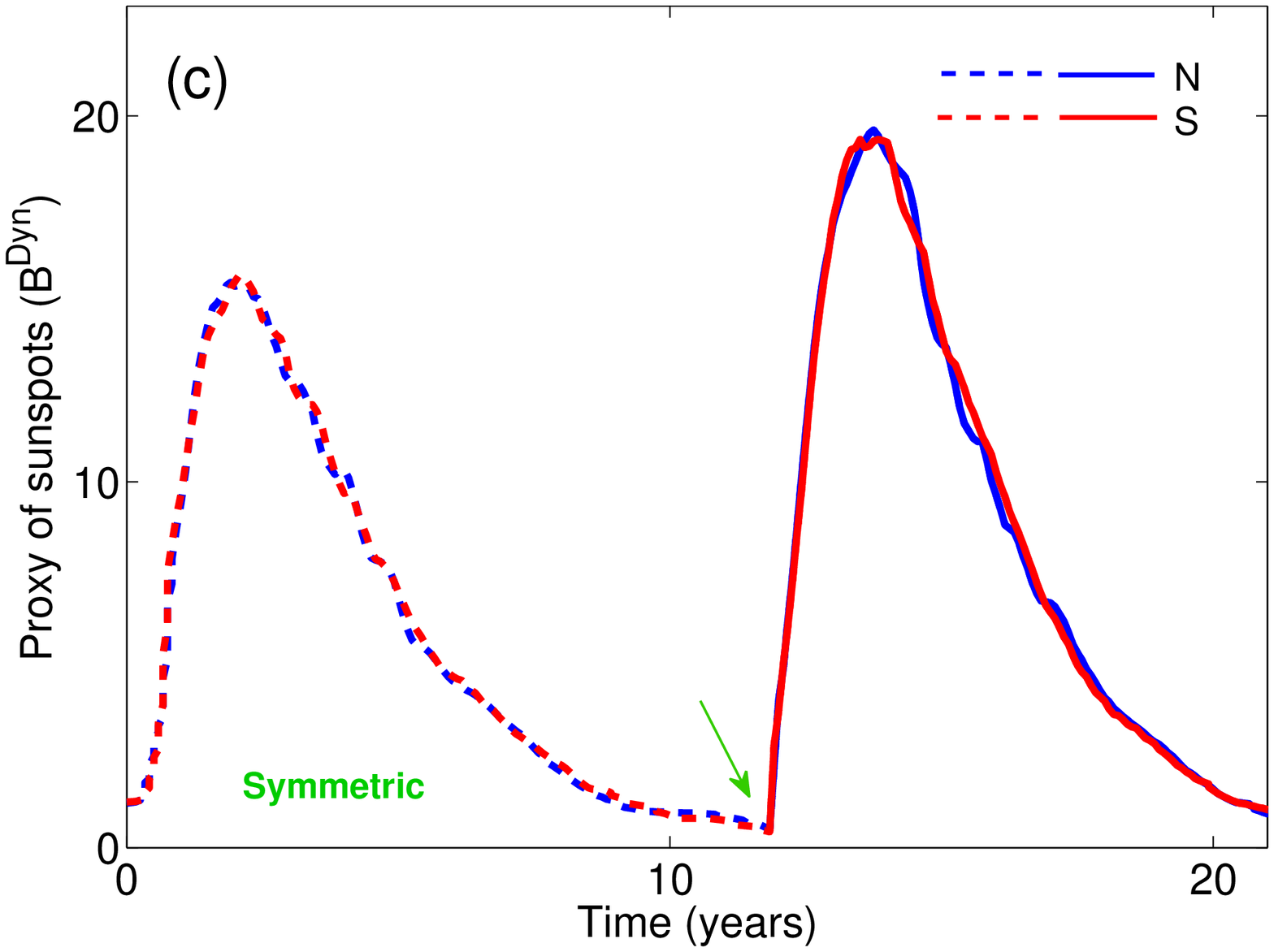}
\includegraphics[height=5.0cm, width=7.5cm]{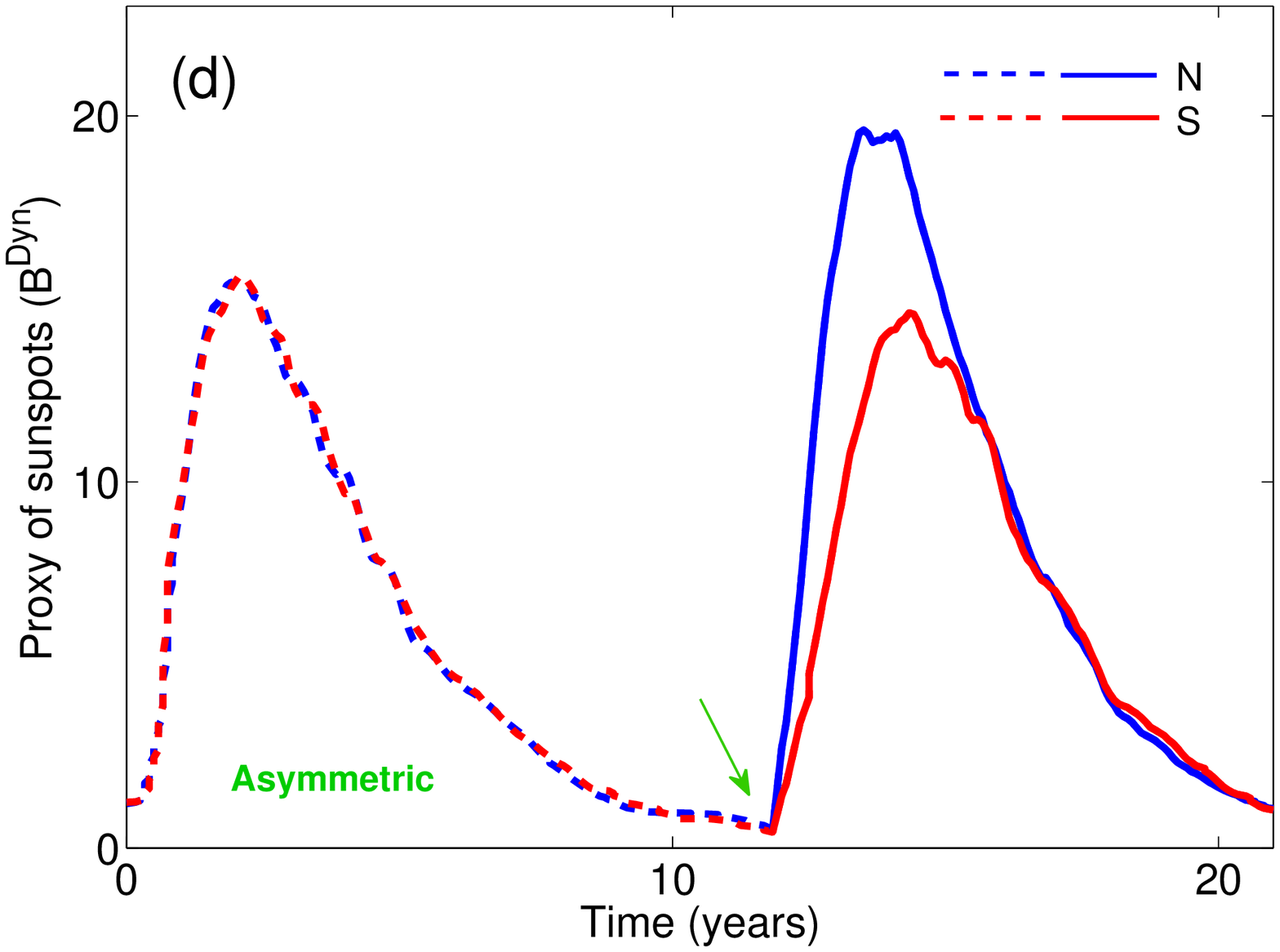}
\caption{(a) represent the vector potentials at cycle minimum on the solar surface as obtained from the SFT and dynamo simulations. (b) depicts two distribution of vector potentials associated with the symmetric and asymmetric cases on the meridional plane. Figures (c) and (d) represent the evolution of B$^{Dyn}$ for the symmetric and asymmetric cases, while the green arrows indicate the timing when outputs obtained from the SFT simulations are assimilated in the dynamo model.} 
\label{dyn_asym}
\end{figure}

Among the four categories discussed in section 4, introducing randomness in tilt angles of active regions results in the maximum hemispheric asymmetry (about 36$\%$). We assimilate the SFT-generated surface magnetic field associated with this case in a dynamo simulation at solar minimum by employing the method outlined above. We represent the vector potentials on the solar surface originated from the SFT (both the symmetric and asymmetric cases) and dynamo simulations as function of latitude in Fig. \ref{dyn_asym}(a); which are further used to obtain the associated $\gamma(\theta)$ functions. From the SFT generated profiles, it's apparent that the northern hemispheric poloidal field (on the surface) for the asymmetric case is slightly stronger than the symmetric one. However, the southern hemisphere is significantly weak in the asymmetric case which is also apparent in the corresponding distribution of vector potentials within the solar convection zone [as depicted in Fig. \ref{dyn_asym}(b)]. Each of these vector potentials is used as an initial condition at solar minimum in a continuous dynamo simulation. 

Fig. \ref{dyn_asym}(c) and (d) depict the time evolution of B$^{Dyn}$ (as described in section 3.2) associated with the northern and southern hemispheres. Following the assimilation of the symmetric vector potential, we find the hemispheric solar activity to be reasonably similar [see the solid blue and red curves in Fig. \ref{dyn_asym}(a)] with an overall increase in strength in both the hemispheres. However, inclusion of the SFT-originated asymmetric vector potential in the dynamo simulation severely affects the solar activity in the southern hemisphere, keeping the northern hemispheric activity almost unaltered [see, Fig. \ref{dyn_asym}(d)]. A detailed analysis of the observational data (in section 2) indicated that the translation of hemispheric asymmetry is better reflected in the total activity associated with a specific hemisphere. We obtain an asymmetry of 21$\%$ (with respect to the symmetric case) by considering the difference between the sum of B$^{Dyn}$ associated with the northern and southern hemispheres. A thorough comparison between hemispheric activity profiles associated with the symmetric and asymmetric cases reveals two aspects: firstly, a reduced amplitude of the poloidal field source originated from the B-L mechanism in the southern hemisphere is well competent to decrease the peak activity during the following cycle in the same hemisphere; and secondly, the timing of peak activity in the corresponding hemisphere can also be shifted. Thus, hemispheric asymmetry is effectively introduced both in the amplitude, as well as in the timing of peak activity during the next cycle which can eventually results in a double peak activity during the following cycle. 

We intriguingly find that the summation $B^{Dyn}$ in the northern hemisphere associated with the asymmetric case decreases by 1$\%$ even if corresponding polar flux increased by 8$\%$ compared to the symmetric case. We speculate this decrement is caused by the coupling between two hemispheres where a weak poloidal field source in the southern hemisphere can effectively reduce the overall activity in the northern hemisphere. 

\section{Reproducing Observed Asymmetry by Data-Driven SFT and Dynamo Simulations}
With the understanding gleaned from the previous sections we now explore the basis of hemispheric asymmetry present in the observed sunspot cycles by conducting numerical simulations with actual observation. We perform a century-scale (1913--2016.75 AD) data-driven SFT simulation starting from solar cycle 15 with a dipolar magnetic field as an initial condition while using the RGO-NOAA/USAF sunspot database which provides information on emergence timing, position and area (thus also the flux) of active regions appearing on the solar surface \cite{2018NatComm...9...5209B}. We note that no hemispheric asymmetry is introduced in this data-driven simulation through transport parameters other than the hemispheric irregularities embedded in the observed sunspot input profile itself. All active regions incorporated in the SFT simulations are assumed to emerge as BMRs and their associated tilt angles is determined by Joy's law with a systematic modification based on cycle amplitude [see equation (15) of \citet{2011A&A...528A..82J}]. We do not introduce any additional randomness in assigning tilt angles to the active regions. 

\begin{figure*}[!h]
\centering
\includegraphics[height=14.0cm, width=15.2cm]{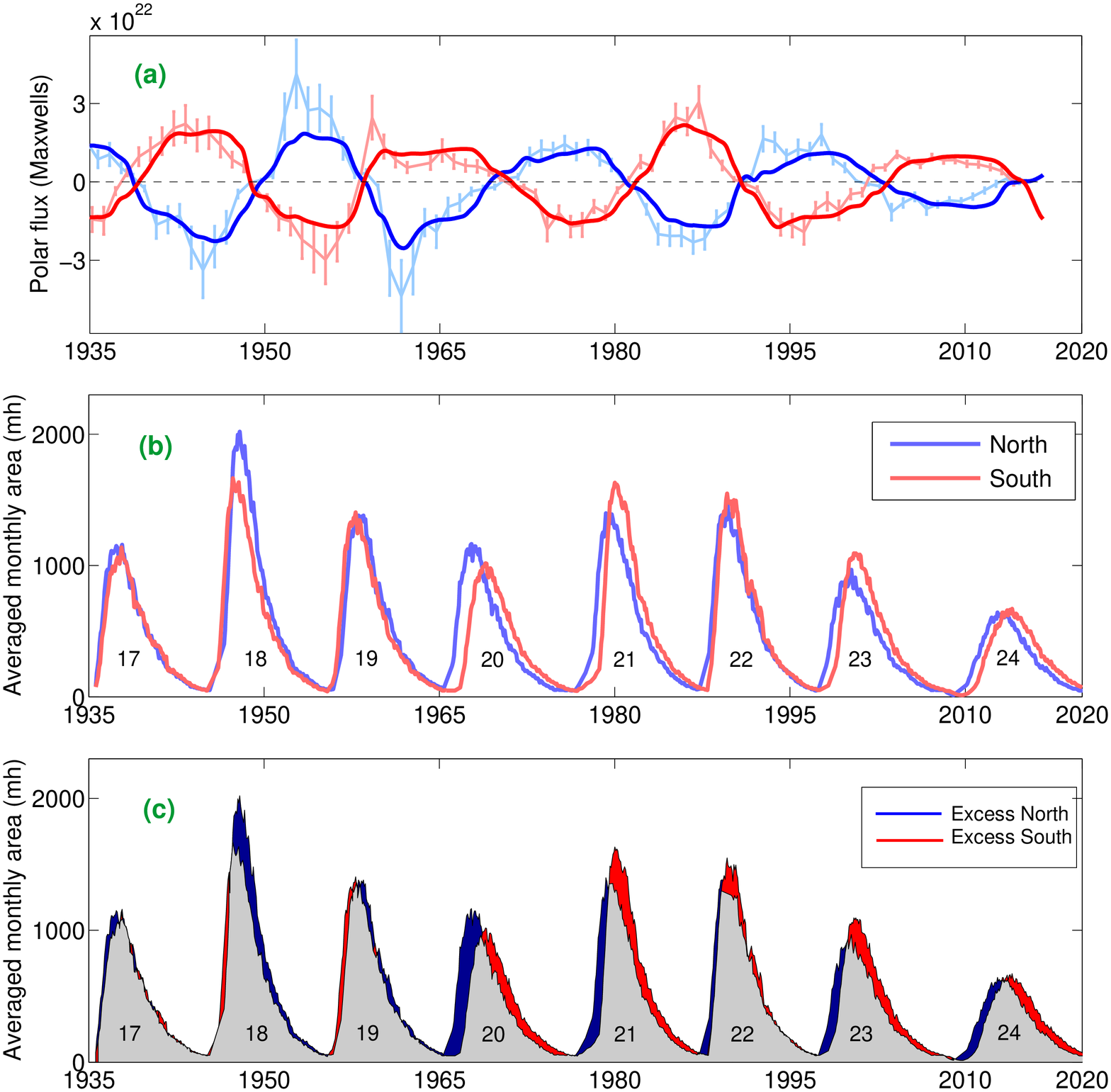}
\includegraphics[height=4.3cm, width=15.2cm]{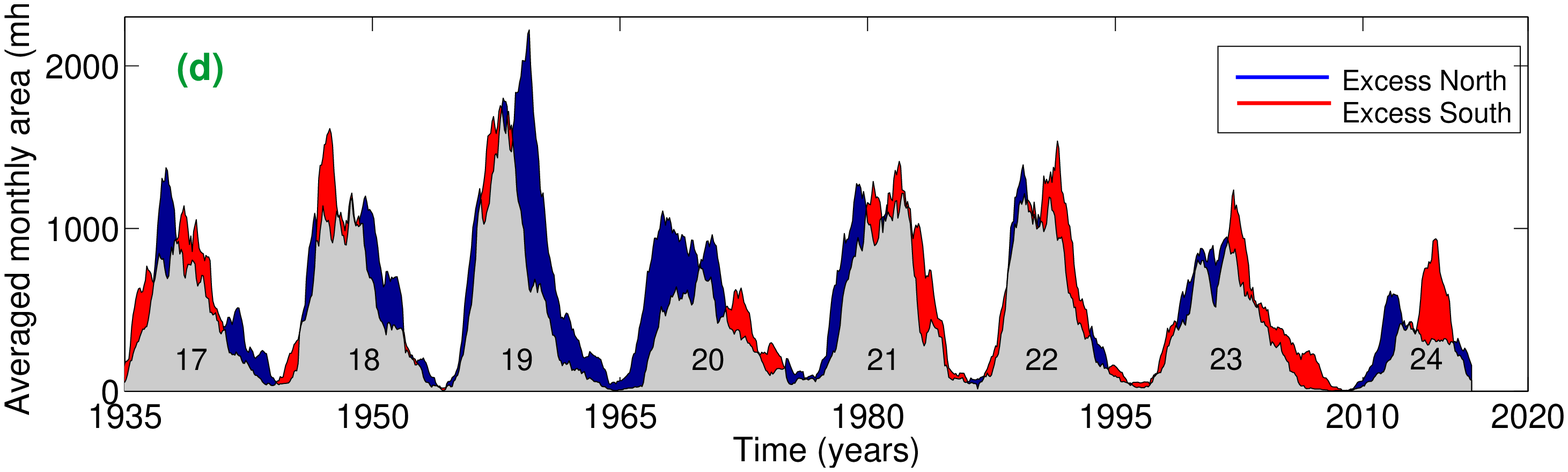}
\caption{The first panel (a) represents the time variation of simulated polar flux in the northern (blue) and southern (red) hemispheres compared against the observed polar flux (north: light blue, south: light red) with error bars. (b) depicts time evolution of hemispheric activities in terms of sunspot area (13 months running averaged) obtained from dynamo simulation. (c) and (d) represent the relative activity (in terms of sunspot area in unit micro-hemispheres) between two hemispheres as obtained from dynamo simulation and observation for solar cycles 17--24, respectively.} 
\label{obs_dyn_cycle}
\end{figure*}

A comparison between the polar flux obtained from SFT simulation and those derived from the MWO polar faculae observation \citep{2012ApJ...753..146M} reveals an overall agreement [see, Fig. \ref{obs_dyn_cycle}(a)] and the simulated polar flux shows a strong correlation (Pearson's correlation coefficient 0.84 with p-value 0.0001) with the observed polar flux at cycle minima. Excluding the northern and southern polar flux data corresponding to cycle 18/19 minimum from our analysis improves the degree of correlation further such that Pearson's correlation coefficient becomes 0.94 with p-value 7.5e-7. Moreover, the north-south asymmetry in polar flux at cycle minimum obtained from the SFT simulations and observations are highly correlated (Pearson's correlation coefficient is 0.91 with p-value 0.0016). In earlier work, \cite{2009RAA.....9..115G} have investigated hemispheric asymmetry by first modeling the poloidal fields at the beginning of solar cycles by utilizing the polar faculae data of the last century and then assimilating them in a dynamo simulation. However, in certain occasions, their results were unable to maintain the basic requirement of one to one correspondence between the hemispheric asymmetry present in polar flux at cycle minimum and the activity of the following cycle (i.e., few data points fall in the 2$^{nd}$ and the 4$^{th}$ quadrants, see Fig. 5 and Fig. 6 in their paper). In contrary, we incorporate the SFT-generated poloidal fields (in terms of vector potential) at every cycle minimum in a continuous dynamo simulation for cycles 17--24 by following the similar method described in section 5. As $B^{Dyn}$ generated by the magnetic buoyancy algorithm serves as a proxy for emerging sunspot flux, we perform calibration between the peak amplitudes of $B^{Dyn}$ and the associated peak in sunspot area for cycle 17--24 to evaluate a constant multiplicative factor. This scaling factor is further utilized to demonstrate the modulation of solar activity (in units of micro-hemispheres) obtained from `SFT-assimilated' dynamo simulations for both the hemispheres [see, Fig. \ref{obs_dyn_cycle}(b)]. A correlation analysis between the amplitudes of peak sunspot area obtain from dynamo simulation and observation gives a Pearson's correlation coefficient of 0.74 with p-value 0.0023 (excluding the data points corresponding to cycle 19.

\begin{figure*}[!h]
\centering
\includegraphics[height=8.2cm, width=18.0cm]{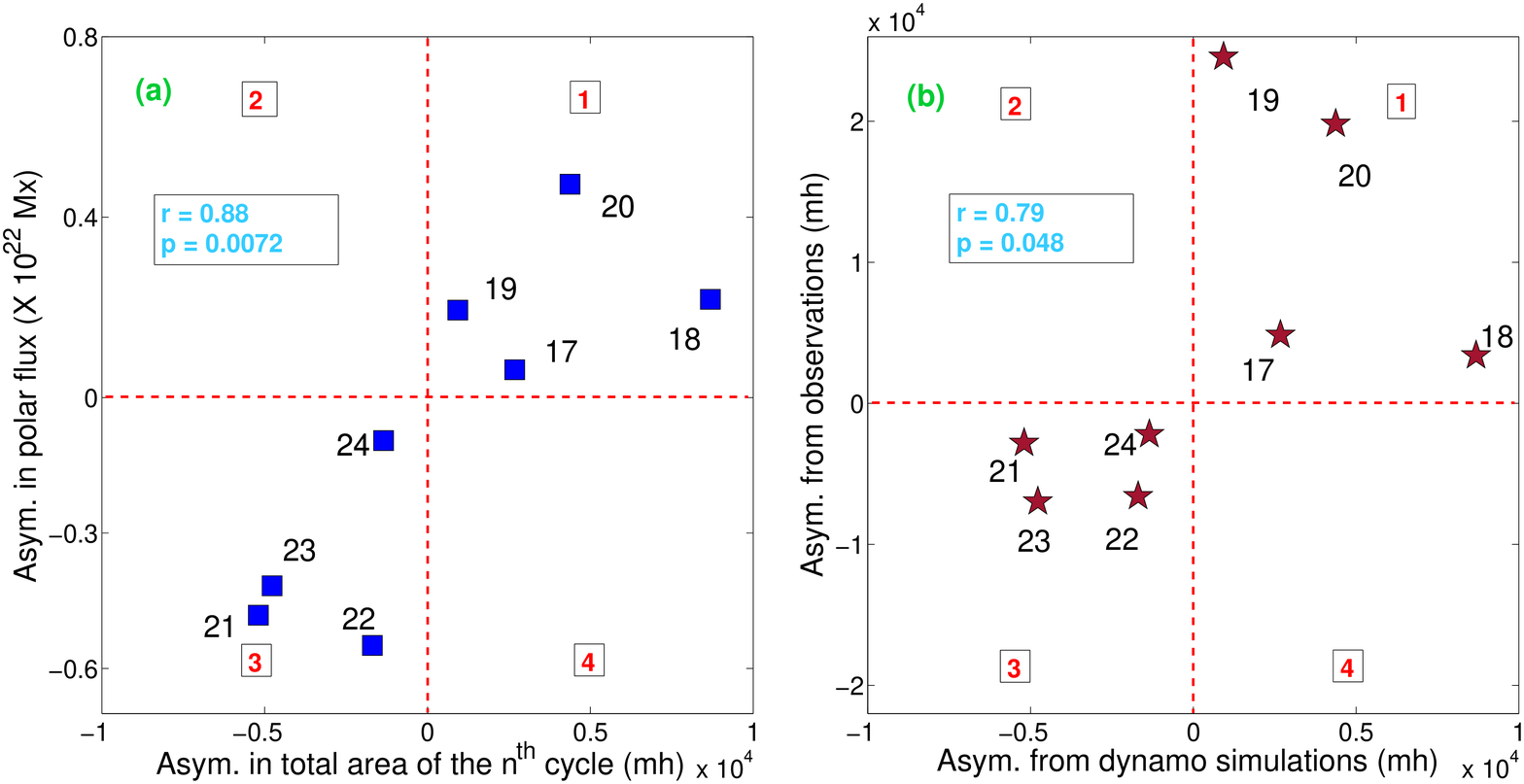}
\caption{(a) represents the hemispheric asymmetry in polar flux (maxwells) during the beginning of the n$^{th}$ cycle (`n' varying from 17 to 24) versus the asymmetry in total sunspot area (in micro-hemispheres) during that cycle. In figure (b), the asymmetry in the total sunspot area during n$^{th}$ cycle as obtained from dynamo simulation and observation are compared. The associated Pearson's correlation coefficients along with the p-values are mentioned.} 
\label{sim_asym}
\end{figure*}

A close inspection of Fig. \ref{obs_dyn_cycle}(b) reveals that north-south asymmetry exists both in the maximum amplitude and epochs of peak magnetic activity in two hemispheres during solar cycles 17--24. Fig. \ref{obs_dyn_cycle}(c) highlights the relative dominance between the northern and the southern hemispheric sunspot area by depicting the time lags present in the rising, peak and declining phases. While comparing with the observed hemispheric asymmetry in sunspot area [see, Fig. \ref{obs_dyn_cycle}(d)], we find that the SFT-assimilated dynamo simulations can reproduce the relative dominance and as well as the phase difference (primarily for cycles 20--24) -- indicating that asymmetry in polar flux at cycle minimum can indeed introduce asymmetry in the following cycle. In Fig. \ref{sim_asym}(a), asymmetry in polar flux at cycle minimum obtained from SFT simulation is plotted against the asymmetry in total hemispheric sunspot area in the following cycle obtained from dynamo simulations. It shows all data points to fall in the 1$^{st}$ and the 3$^{rd}$ quadrants -- satisfying the essential condition of positive correlation between these two quantities. We find Pearson's correlation coefficient to be 0.88 with p-value 0.0072. A comparison between the observation and the north-south asymmetry calculated from total sunspot area derived from dynamo simulation [see Fig. \ref{sim_asym}(b)] shows that our simulations are successful in preserving the relative hemispheric dominance while the degree of correlation is quite substantial (Spearman's rank correlation coefficient is 0.79 with p-value 0.048, excluding cycle 19 from our analysis). We note that even though results generated from simulations capture the nature of relative time difference in the epochs of peak activity for cycles 20--24 (i.e., which hemisphere peaks earlier in a cycle), the amplitude of phase lag is not same as the seen in observation. 

We attribute the discrepancies between the simulated results and observations [see, Fig. \ref{obs_dyn_cycle}(a) and (b)] to different aspects of our assumptions used in dynamo simulations. In this work, we assume the origin of hemispheric asymmetry in the poloidal field source to be solely the B-L mechanism while considering the amplitude of the mean field $\alpha$-effect to remain constant throughout the simulations (i.e., free from any hemispheric irregularities). Additionally, earlier studies \citep{2003ApJ...589..665H,2004ESASP.559..241N,2008ApJ...673..544Y} have found that the speed of equator-ward meridional flow controls the duration of the solar cycle along with the timing of peak activity to some extent which we have not accounted in this current work. Thus, any profound hemispheric irregularities in meridional circulation speed can induce a relative phase difference in the rising, peak and declining epochs of the northern and the southern hemispheric magnetic activities. 

\section{Discussions and Conclusions}

A detailed analysis of the observational data shows a strong positive correlation between the polar flux during cycle minimum and the peak activity of the following cycle in both the northern and the southern hemispheres -- which is in agreement with earlier studies. Moreover, the north-south asymmetry in polar flux at cycle minimum is strongly correlated (the relation being non-linear) with the asymmetry in the overall hemispheric activity during the following cycle. Additionally, we observe that the polar flux can attain its maximum amplitude much before the solar minimum. However, the maximum amplitude of the hemispheric polar flux and the associated timing ($t_N$ or $t_S$) have no impact on the following cycle regarding the timing and magnitude of peak activity in the corresponding hemisphere. This observational feature establishes the amplitude of polar flux at the solar minimum as a better precursor for forecasting the strength of the following cycle -- strongly supporting the concept utilized in various studies of solar cycle prediction \citep{2016ApJ...823L..22C,2016JGRA..12110744H}.

We have performed multiple SFT simulations by incorporating diverse irregularities associated with the B-L mechanism in the northern and the southern hemispheres. We consider three aspects of hemispheric asymmetry associated with the sunspots emerging on the solar surface. A time-gap of 2.5 years between the peak hemispheric activity results in a time difference of approximately one year in the reversal timing of the polar flux in two hemispheres. Given the initial polar field strength (at the beginning of the cycle) and the total flux associated with the sunspots are equal in both the hemispheres, the final polar field strength during cycle minimum remains unaffected by the imposed time-gap. Since the cancellation of the older polar field and development of the new one materialize through a slow process, a temporal discrepancy in the peak hemispheric activity does not induce any asymmetry in the final polar field strength. 

An imbalance in the total magnetic flux associated with sunspots in the northern and southern hemispheres modifies the final polar flux in both the hemispheres -- such that, a 70$\%$ asymmetry in sunspot associated flux induces only 14$\%$ hemispheric asymmetry in the final polar flux strength. In the absence of sufficient leading polarity spots in the northern hemisphere, magnetic flux from both the leading and following polarities advects towards the south pole, effectively weakening the polar flux in the more active hemisphere. Moreover, magnetic flux from the leading polarity may cross the equator and traverse towards the pole of the other hemisphere in case of an extreme asymmetry where almost no spot appears on the other hemisphere. Incorporating randomness in the tilt angles of active regions produces the maximum asymmetry in the final amplitude of the polar flux (as high as 36$\%$). The only significant transport parameter through which hemispheric irregularities can be introduced in the B-L mechanism is the meridional circulation. However, we find that a north-south asymmetry of $\pm$50$\%$ in the peak amplitude of meridional flow generates only 3$\%$ asymmetry in the polar field strength at cycle minimum -- the reason of which has been explained in the context of cross-equatorial flux cancellation between the leading polarity spots in two hemispheres. In summary (refer to Table 1), irregularities present in tilt angle and areal coverage of sunspots emerged on the solar surface during a cycle can induce significant hemispheric asymmetry in both the timing of reversal and final amplitude of the polar field whereas we can neglect the contribution from relative discrepancy in the peak flow speed of meridional circulation between two hemispheres.

We explore the extent to which the north-south asymmetry present in the polar field at solar minimum is translated to the hemispheric activity in the succeeding cycle. We find that a critically weak poloidal field source in the southern hemisphere (where $\Phi_p^{S}$ is 36$\%$ weaker than $\Phi_p^{N}$) strongly modulates the amplitude and timing of peak activity in the same hemisphere during the following magnetic cycle. This eventually introduces an asymmetry in the overall hemispheric activity (i.e., a summation of $B^{Dyn}$) and a profound time gap between the occurrence of peak activity in two hemispheres. Additionally, a hemispheric coupling reduces the overall strength in the northern hemisphere, effectively, decreasing the magnitude of asymmetry. However, we surmise that the degree of this coupling is subjected to the profiles associated with the parameters used in dynamo simulations. 

Analyzing the results obtained from the century-scale data-driven SFT and dynamo simulations we establish that hemispheric asymmetry present in the poloidal field source at cycle minimum originated from the B-L mechanism is capable of inducing significant asymmetry in hemispheric sunspot activity in the following solar cycle. However, other factors like fluctuations in mean field $\alpha$-effect and hemispheric irregularities in the meridional circulation flow speed can also play a crucial role in reproducing the exact hemispheric asymmetry as observed in the past sunspot cycles. We speculate that a detailed analysis with diverse configurations of the initial poloidal field source and an extended parameter space study along with different levels of fluctuations can reveal essential aspects of the underlying physics involved in the dynamo mechanism -- which we plan to address in a following work.

In summary, our analyses elaborate intricate characteristics of the B-L mechanism with a primary focus on the key elements causing hemispheric asymmetry in the large-scale polar field of the Sun. We demonstrate the importance of cross-equatorial cancellation of magnetic flux among the leading polarity spots associated with two hemispheres for the development of the new polar field following a polarity reversal at solar maximum. Lastly, by assimilating synthetic and observed data-driven SFT results in dynamo simulations, we have illustrated that an asymmetric poloidal field at solar minimum is capable of introducing notable asymmetry in both the amplitude and phase of hemispheric activities in the succeeding cycle -- indicating that such asymmetry can be a potential basis for generation of double peak cycles. 

\begin{acknowledgements}
The author acknowledges utilization of data from the Royal Greenwich Observatory/USAF-NOAA active region database compiled by David H. Hathaway \url{(https://solarscience.msfc.nasa.gov/greenwch.shtml)}. MWO calibrated polar faculae data were downloaded from the solar dynamo database maintained by Andr\'{e}s Mu{\~n}oz-Jaramillo \url{(https://dataverse.harvard.edu/dataverse/solardynamo)}. The author acknowledge funding by CEFIPRA/IFCPAR through grant 5004-1. The author is thankful to Prof. Dibyendu Nandy for fruitful discussions and suggestions which helped to improve the quality of this work.
\end{acknowledgements}

\bibliographystyle{aa} 

\newpage
 
\vspace{10cm}
\begin{table}[!hb]
\begin{center}
\begin{tabular}{ |p{2.9cm}|p{2.0cm}|l|l|l|l| }
\hline
\multicolumn{3}{ |c| }{{ \bf Table 1}: Hemispheric Asymmetry} \\
\hline
Introduced in & {Amount} & Change in  \\ 
{} & {} & Final Polar Flux\\
\hline
&  &  \\ 
Scatter in Tilt Angle & -- & 36$\%$  \\ 
&  &  \\ 
\hline
&  &  \\ 
Sunspot Flux & 70$\%$ & 14$\%$  \\ 
&  &  \\ 
\hline 
&  &  \\ 
Peak Flow Speed of Meridional Circulation & 50$\%$ & 3$\%$ \\
&  &  \\ 
\hline
&  &  \\ 
Phase-lag in Peak Activity & 2.5 years & none\\
&  &  \\ 
\hline
\end{tabular}
\vspace*{0.1cm}
\end{center}
\end{table}

\end{document}